\documentclass[aps,preprint,showpacs,preprintnumbers,amsmath,amssymb]{revtex4}

\usepackage{graphicx}
\usepackage{subfigure}
\usepackage{epsfig}

\newcounter{saveeqn}

\begin{document}

\title{Exact ground state of finite Bose-Einstein condensates on a ring}
\author{Kaspar \ Sakmann\footnote{Corresponding author, 
E-mail: kaspar@tc.pci.uni-heidelberg.de},\\ \vspace{0.15cm} Alexej I.\ Streltsov\footnote{E-mail: alexej@tc.pci.uni-heidelberg.de}, 
Ofir E. Alon\footnote{E-mail: ofir@tc.pci.uni-heidelberg.de} and Lorenz S.\ Cederbaum\footnote{E-mail: lorenz.cederbaum@urz.uni-heidelberg.de}}
%\vspace{1cm}
%\author{O. E.\ Alon, A. I.\ Streltsov and L. S.\ Cederbaum}
%\vspace{0.15cm}

%\\ Alexej I.\ Streltsov, Ofir E. Alon,  Lorenz S.\ Cederbaum} 
%\author{K.\ Sakmann\footnote{E-mail: kaspar@tc.pci.uni-heidelberg.de},\\ O. E.\ Alon, A. I.\ Streltsov and L. S.\ Cederbaum} 
%\author{Ofir E.\ Alon} 
%\author{Alexej I.\ Streltsov} 
%\author{Lorenz S.\ Cederbaum} 
\affiliation{Theoretische Chemie, Universit\"at Heidelberg, 69120 Heidelberg, Germany}
\date{\today}

\begin{abstract}
The exact ground state of the many-body Schr\"odinger equation for $N$ bosons on a one-dimensional 
ring interacting via pairwise $\delta$-function interaction is presented for up to fifty particles. 
The solutions are obtained by solving Lieb and Liniger's system of coupled transcendental equations for finite $N$. 
The ground state energies for repulsive
and attractive interaction are shown to be smoothly connected at the point of zero interaction strength, 
implying that the \emph{Bethe-ansatz} can be used also for attractive interaction for all cases studied.
For repulsive interaction the exact energies are compared to 
(i) Lieb and Liniger's thermodynamic limit solution and 
(ii) the Tonks-Girardeau gas limit.
It is found that the energy of the thermodynamic limit solution can differ substantially from that of
the exact solution for finite $N$ when the interaction is weak or when $N$ is small.
A simple relation between the Tonks-Girardeau gas limit and the solution for finite 
interaction strength is revealed. For attractive interaction we find that the true ground state energy 
is given to a good approximation by the energy of the system of 
$N$ attractive bosons on an infinite line, provided the interaction is stronger 
than the critical interaction strength of mean-field theory.
\end{abstract}


\maketitle
\section{Introduction}
The recent experimental realization of quasi one-dimensional Bose-Einstein condensates 
\cite{expONED1,expONED2,expONED3,expONED4,expONED5,expONED6,expONED7} has made
the theoretical description of these systems a very active area of research.
In this context, the underlying equation is the many-body Schr\"odinger equation for $N$ particles subject to two-body
$\delta$-function interaction and possibly an external potential
\cite{Olshanii}.
In general, solving this equation exactly is very difficult, and it is almost
inevitable to introduce approximations.
The most commonly used approximation for Bose-Einstein condensates is the 
Gross-Pitaevskii (GP) approximation \cite{GP1,GP2}. 
The GP approximation is a mean-field approximation that results in a comparatively simple nonlinear equation.
This equation can be solved numerically and 
explicit analytical solutions have been found for some cases, see \cite{Reinhardtattr, Reinhardtrep, Ueda} 
and references therein.
Despite its great success in the description of early experiments, 
the GP approximation suffers from various shortcomings. For instance, the solutions of the GP equation 
may not possess the symmetry of the Hamiltonian of the problem 
\cite{Calogero,Reinhardtattr, Reinhardtrep, Ueda}. 
More recently, other approximations have been developed
to overcome these difficulties, see \cite{overview,BMF1,CCI,CCI2}.

Exactly solvable one-dimensional models are of interest by themselves
and can be considered as a research discipline of its own, see \cite{1Da,1Db} for an overview.
Moreover, they provide an invaluable testing ground for approximative methods.
In fact, this is often the motivation for the research in this field. 
However, from an experimental point of view the assumption 
of one-dimensionality seems far-fetched at first sight.
Therefore, it is very  exciting to see the experimental realization of some
of these model systems within reach.

A very prominent exactly solvable model is Lieb and Liniger's system of $N$
spinless point-like bosons on a one-dimensional ring interacting via pairwise 
$\delta$-function interaction \cite{LiebLiniger}. 
This is also the subject of the present work, and the proper definition of the problem 
is given in the next section.
Lieb and Liniger's model is a generalization of Girardeau's gas of impenetrable bosons \cite{Girardeau}
to finite interaction strength.
The impenetrable boson gas is also known as the Tonks-Girardeau (TG) gas, 
thereby including also the name of the inventor of the classical hard sphere gas \cite{Tonks}.
The TG gas is not less prominent than its finite interaction counterpart,
and even though both models are now more than forty years old, 
they are still the subject of ongoing research, see \cite{pres1,pres2,pres3,pres4,pres5}.

In their ground-breaking work Lieb and Liniger derived a system of $N-1$ coupled transcendental equations
that determine the exact $N$-particle ground state of the problem. 
Lieb and Liniger solved this system explicitly for two particles, but then passed to the thermodynamic 
limit of the system which exists for repulsive interaction only.
Surprisingly, in this limit the whole system of coupled transcendental equations can be approximated
by a single Fredholm integral equation of the second kind.
Lieb and Liniger solved this integral equation already in their initial work.
Moreover, they proved that it has an analytic solution for any
interaction strength truly greater than zero.
For weak interaction, Bogoliubov's perturbation theory agrees well with 
this solution and for strong interaction its energy converges to that 
of the TG gas in the thermodynamic limit \cite{Girardeau}.
In a subsequent paper Lieb also derived the excitation spectrum of 
the thermodynamic limit solution \cite{Lieb}.
Lieb and Liniger's thermodynamic limit solution has also been used to describe interacting bosons 
in more general one-dimensional trapping potentials than just a one dimensional ring.
For example, by assuming that the thermodynamic limit approximation is locally valid and by employing a 
hydrodynamic approach, bosons in cigar shaped traps have been described \cite{cigar}.

However, the set of coupled transcendental equations which yield the exact solution 
of the \emph{finite} $N$ problem was not solved for $N>2$ until 1998, neither for 
repulsive nor for attractive interaction. In 1998 Muga and Snider derived the whole spectrum 
of the three-particle problem \cite{MugaSnider} for attractive and repulsive interaction.
Still, up to date there is no exact solution of the problem for $N>3$.
In the present work we would like to fill in this gap. 
We calculate the ground state for up to fifty particles.
For repulsive interaction even for fifty particles the thermodynamic limit solution can deviate from the
solution for finite $N$ by as much as two percent, as we shall show. 
For attractive interaction we reveal a close relation to the system of $N$ interacting 
bosons on an infinite line, see \cite{Calogero} and references therein.

It has been proven that for repulsive interaction the wave functions of all 
states are of the \emph{Bethe-ansatz} type \cite{YangYang,Dorlas}.
This is not the case for attractive interactions \cite{LiebLiniger, MugaSnider}.
However, Muga and Snider have shown that for three attractive bosons \cite{MugaSnider} 
a complete set of states can probably be \emph{derived} from a \emph{Bethe-ansatz}.
We will clarify this at a later stage. 
In the present work we give further support 
to the hypothesis that all states may be derived from a \emph{Bethe-ansatz} for 
all particle numbers and all interaction strengths.

As we have shown recently, the problem of $N$ bosons on a ring defies any accurate description by using 
direct diagonalization techniques, except for very weak interaction \cite{CCI}.
In the present work, we present the exact ground state solution of the system of $N$ bosons on a one-dimensional ring
by solving Lieb and Liniger's system of coupled transcendental equations for up to fifty particles.
We treat repulsive and attractive interactions alike and compare the results 
with other limiting cases of this system.
Our approach can easily be extended to any particle number and also to excited states.

The plan of the paper is as follows. In section \ref{Theory} we define the problem, review the 
derivation of Lieb and Liniger's system of
coupled transcendental equations and derive results for weak attractive and repulsive interaction, by using first order perturbation theory.
Section \ref{NumericalSolution} addresses the problems related to the numerical solution of the set of coupled transcendental equations.
The attractive case proved to be particularly delicate.
We show that the \emph{Bethe-ansatz} gives solutions for attractive and repulsive interaction that are continuously
connected to each other at the point of zero interaction for all particle numbers under consideration.
In section \ref{resultsonrepulsivebosons} we present the exact energies for repulsive interaction for up to fifty particles.
We compare these energies with the energies of first order perturbation theory, 
the TG gas and the thermodynamic limit. Furthermore, we present the differences of these limits to the finite $N$ solution
and reveal a surprisingly simple relation between the TG limit and our solution.
We give explicit limits on the minimal number of particles and the size of the ring for the thermodynamic limit approximation to be 
reasonably accurate.
In section \ref{resultsonattractivebosons}, the results for attractive interaction are presented.
The exact ground state energies are found to converge to the energies of the one-dimensional problem of $N$ attractive bosons on
an infinite line, see \cite{Calogero} and references therein.
We give a simple explanation for this behaviour in a mean-field picture and 
show that this energy is approached in the limit of very strong attractive interaction.
To be concrete, we also calculate the minimal number of particles which is 
necessary such that the system on a ring of finite size 
can be approximated by the system on an infinite line.
Section \ref{Conclusions} contains a summary of our results and a discussion of open questions. 

\section{Theory \label{Theory}}
\subsection{The Schr\"odinger equation of the problem}

Our starting point is the stationary Schr\"odinger equation for $N$
bosons in one dimension subject to two-body $\delta$-function interaction
and periodic boundary conditions:
% We put $\hbar=1$ ,$m=\frac{1}{2}$
%and obtain:
\begin{equation}
-\frac{\hbar^2}{2m}\sum_{i=1}^{N}\frac{\partial^{2}\psi}{\partial y_{i}^{2}}+2 \tilde c\sum_{i<j}\delta(y_{i}-y_{j})\psi= \tilde E_l(\tilde c)\psi,
\label{DimRingEq}
\end{equation}
where the $y_{i}$ are the particle coordinates $0\le y_{i}\le l$
and 
\begin{equation}
\psi(y_{1},...,y_{i},...,y_{N})=\psi(y_{1},...,y_{i}+l,...,y_{N}),
\end{equation}
for $i=1,...,N$. We divide equation (\ref{DimRingEq}) by $\frac{\hbar^2}{2m}$ 
and change from the dimensional to dimensionless coordinates
$x_{i}=\frac{y_{i}}{l}L$, where $L$ is the new dimensionless length of the ring. 
The combined effect of these changes results in the equation 
\begin{equation}
-{\displaystyle \sum_{i=1}^{N}\frac{\partial^{2}\psi}{\partial x_{i}^{2}}}+2c\sum_{i<j}^{N}\delta(x_{i}-x_{j})\psi=E_L(c) \psi, 
\label{DimlessRingEq}
\end{equation}
where now $0 \le x_{i}\le L$.
The relation between the dimensionless and the dimensional quantities is given by the equations
\begin{equation}
y_{i}=\frac{x_{i}}{L}l,
\end{equation}
\begin{equation}
\tilde E_l(\tilde c) = \frac{\hbar^2}{2m} \frac{E_L(c)L^2}{l^2}, 
\label{Etilde}
\end{equation}
\begin{equation}
\tilde c = \frac{\hbar^2}{2m} \frac{cL}{l},
\label{ctilde}
\end{equation}
From (\ref{Etilde}) and (\ref{ctilde}) it follows that the relation between 
the dimensionless energies of two rings, one of length $L$ and one of length $L^{\prime}$, 
is given by the equation
\begin {equation}
E_{L^{\prime}}(c)=\frac{E_L(\frac{L^{\prime}}{L} c)}{\frac{{L^{\prime}}^2}{L^2}}.
\label{scaleEnergy}
\end{equation}
Attractive interactions are described 
by $c<0$ and repulsive by $c>0$.

\subsection{The \emph{Bethe-ansatz} wave function}
The $\delta$-function potential in (\ref{DimlessRingEq}) is equivalent to a jump in 
the derivative of the wave function, wherever two particles touch \cite{LiebLiniger}:
\begin{equation}
\left(\frac{\partial\psi}{\partial x_{j}}-\frac{\partial\psi}{\partial x_{k}}\right)_{x_{j}=x_{k}+}-\left(\frac{\partial\psi}{\partial x_{k}}-\frac{\partial\psi}{\partial x_{j}}\right)_{x_{j}=x_{k}-}=2c\psi|_{x_{j}=x_{k}}.
\label{jumpcondition}
\end{equation}
As long as $c<\infty$ the interaction potential allows the particles
to go past each other, and any particle can be anywhere in coordinate
space. However, since the particles are identical bosons the knowledge of the wave function 
in the "primary" region 
\begin{equation}
R_{p}:0\le x_{1}\le x_{2}\le...\le x_{N}\le L,
\end{equation} namely
\begin{equation}
\psi(x_{1}\le x_{2}\le...\le x_{N})
\label{WFRp}
\end{equation}
contains the full information.  
It has to be stressed that the wave function in $R_{p}$
does \emph{not} have to be symmetric.
Once the wave function in $R_{p}$ is known, the wave function
in any other region corresponding to a different ordering of the coordinates
is obtained simply by interchanging the particle labels in (\ref{WFRp}).
This ensures the symmetry of the total wave function under particle exchange in
the unrestricted coordinate space. 
In the region $R_{p}$, equations (\ref{DimlessRingEq}) and (\ref{jumpcondition}) become
\begin{equation}
-\sum_{i=1}^{N}\frac{\partial^{2}}{\partial x_{i}^{2}}\psi=E_{L}(c)\psi, \label{exact energy}
\end{equation}
for $x_{i}\ne x_{j}$ and
\begin{equation}
\left(\frac{\partial}{\partial x_{j+1}}-\frac{\partial}{\partial x_{j}}\right)\psi|_{x_{j+1}=x_{j}}=c\psi|_{x_{j+1}=x_{j}}.\label{jump cond R_p}
\end{equation}
Since periodic boundary conditions are used, a displacement of $L$ in
any of the coordinates leaves the wave function unchanged. Applying
a displacement of $L$ to the wave function in $R_{p}$ yields
\begin{equation}
\psi(0,x_{2},...,x_{N})=\psi(x_{2},x_{3},...,x_{N-1},L)\label{periodic R_p}
\end{equation}
and for the derivatives 
\begin{equation}
\frac{\partial}{\partial x}\psi(x,x_{2},...,x_{N})|_{x=0}=\frac{\partial}{\partial x}\psi(x_{2},x_{3},...,x_{N},x)|_{x=L}.
\label{periodic deriv R_p}
\end{equation}
Two particles only interact when they are at the same
point in space. This fact and (\ref{exact energy}) motivate the idea
that the solution might be a product of plane waves.
In fact, the \emph{Bethe-ansatz} is just a more general version of this
idea, namely a superposition of products of plane waves. 
It was first applied to spin chains \cite{Bethe}, but has
been successfully used to solve numerous other one-dimensional problems exactly \cite{1Db}.
The \emph{Bethe-ansatz} wave function for this problem is \cite{LiebLiniger}
\begin{equation}
\psi(x_{1}\le x_{2}\le...\le x_{N})=\sum_{P}a(P)\, P\, exp(i\sum_{j=1}^{N}k_{j}x_{j}),\label{Bethe ansatz}
\end{equation} 
where the sum runs over all permutations $P$ of the $\{ x_{j}\},$
and the $a(P)$ are coefficients which are determined by the rule given below. 
It has been proven that the \emph{Bethe-ansatz} gives all solutions 
of the problem for repulsive interaction  \cite{Dorlas,YangYang}.
For the \emph{Bethe-ansatz} to be valid, all $k_{i}$ must be different from one another if $c \ne 0$,
otherwise $\psi$ vanishes identically by means of (\ref{jump cond R_p}).
Only if $c=0$, the equality of two $k_j$ does not imply a vanishing \emph{Bethe-ansatz} wave function.
However, there are certain critical $c$ values at which the equality of two $k_j$ \emph{does} occur \cite{MugaSnider,LiebLiniger}, implying
that the form of the \emph{Bethe-ansatz} wave function is no longer valid.
This only reflects the fact that the normalization constant has not been included in the definition of the \emph{Bethe-ansatz} wave function.
The proper normalized wave function can be obtained by  using the rule of l'H\^ospital \cite{MugaSnider}.
As Muga and Snider have shown, for three particles all \emph{Bethe-ansatz} solutions are continuously connected
in $k$-space at $c=0$.
It is therefore very likely that for three particles a complete set of states can be derived from the \emph{Bethe-ansatz}
in the sense that l'H\^ospital's rule is applied to obtain the wave function from the \emph{Bethe-ansatz} at the critical $c$-values.
As we shall show in section \ref{NumericalSolution}, the \emph{Bethe-ansatz} 
provides ground state solutions that are continuously connected 
in $k$-space at $c=0$, at least for as many as fifty particles. 
Therefore, we suspect that a complete set of states can be derived for 
all particle numbers from a \emph{Bethe-ansatz} wave function in the sense described above.

\subsection{Lieb and Liniger's transcendental equations}
In the following we review the derivation of Lieb and Liniger's system of coupled transcendental 
equations as far as it is indispensable for our needs.
However, we adopt Muga and Snider's approach and  notation, since it allows to treat 
the repulsive and the attractive case in a coherent fashion.

The coefficients $a(P)$ in (\ref{Bethe ansatz}) are obtained
by the following rule \cite{LiebLiniger}: For $P=I$, the identity, set $a(I)=1$.
For any other permutation decompose $P$ into transpositions.
For every transposition of the particles $j$ and $l$ write down
a factor of $-e^{i\theta_{jl}}$.
For example, if \begin{equation}
P=\left(\begin{array}{c}
123\\
321\end{array}\right)=\left(\begin{array}{c}
123\\
132\end{array}\right)\left(\begin{array}{c}
132\\
312\end{array}\right)\left(\begin{array}{c}
312\\
321\end{array}\right),\end{equation}
then one obtains the result 
\begin{equation}
a(P)=-e^{i(\theta_{32}+\theta_{31}+\theta_{21})},
\end{equation}
where 
\begin{equation}
e^{i\theta_{jl}}=\frac{c-i(k_{j}-k_{l})}{c+i(k_{j}-k_{l})}
\label{defTheta_jl}
\end{equation}
or equivalently
\begin{equation}
\theta_{jl}=i\, log\left[\frac{c+i(k_{j}-k_{l})}{c-i(k_{j}-k_{l})}\right]=-2\, arctan\left(\frac{k_{j}-k_{l}}{c}\right),
\label{theta_jl}
\end{equation}
where the branch of the logarithm or the $arctan$ has not been specified yet. 
A more detailed description of how the coefficients $a(P)$
are obtained can be found elsewhere \cite{LiebLiniger}.

The condition (\ref{jump cond R_p}) determines only
the form of the wave function. The allowed values for the set $\{ k_j\}$
have to be determined by the periodicity conditions (\ref{periodic R_p})
and (\ref{periodic deriv R_p}). In fact, these equations are equivalent to \cite{LiebLiniger}:
\begin{equation}
(-1)^{N-1}e^{-ik_{j}L}=exp(i\sum_{s=1}^{N}\theta_{sj}),\,\,\,\, j=1,2,...,N,
\label{defining eq}
\end{equation}
where $\theta_{jj}=0$. 
This set of coupled transcendental equations determines the wave vectors $k_j$. 
However, in the present form it is still very cumbersome to work with, 
and it is customary to introduce new variables which exploit the symmetries of the problem.
By solving (\ref{defining eq}) for $k_{j}L$,
one arrives at 
\begin{equation}
k_{j}L=2\pi m_{j}-\sum_{s=1}^{N}\theta_{sj},\,\,\,\, j=1,2,...,N,
\label{k_j}
\end{equation}
for some integers $\{ m_{j}\}$. 
The combination of (\ref{k_j}) and (\ref{theta_jl}) results in Lieb and Liniger's system of $N-1$ coupled transcendental equations
for the differences between the $\{k_j\}$:
\begin{equation}
(k_{j+1}-k_{j})L=i\, Log\left[\frac{\prod_{l=1}^{N} \frac{c+i(k_{l}-k_{j})}{c-i(k_{l}-k_{j})}}{\prod_{m=1}^{N}\frac{c+i(k_{m}-k_{j+1})}{c-i(k_{m}-k_{j+1})}}\right]+2\pi n_{j},\,\,\, j=1,...,N-1,
\label{DEFINING EQS}
\end{equation}
where now the principal part of the logarithm is taken and the integers ${n_j}$ are discussed further below.
Since the principal part of the logarithm is taken in (\ref{DEFINING EQS}) and not in (\ref{k_j}), $n_j$ 
is not necessarily equal to $m_{j+1}-m_{j}$.
Every state is uniquely defined by the set $\{k_{j}\}$. 
The order of these is unimportant, since
the particles are identical bosons. For real $\{k_{j}\}$ they can be ordered
such that 
\begin{equation}
k_{1}\le k_{2}\le ...\le k_{N}
\label{order k_j}
\end{equation}
which can be satisfied by choosing
\begin{equation}
-2\pi< \Re(\theta_{jl})\le0,
\label{range theta_jl}
\end{equation}
if $j>l$. The choice (\ref{range theta_jl}) implies a one-to-one correspondence between the $\theta_{jl}$ and the set $\{k_j\}$, 
provided (\ref{DEFINING EQS}) have a unique solution for every set $\{m_j\}$, which can be justified \cite{LiebLiniger}.   
When $c$ varies continuously from $-\infty$ to $+\infty$, 
the arguments of the logarithms in the $\theta_{jl}$ also vary continuously.
When the argument of any of the $\theta_{jl}$ arrives at the discontinuity of the branch cut,
the $\theta_{jl}$ are continued analytically to compensate for the branch cut discontinuity.

By taking the product of all $N$ equations (\ref{defining eq}) one arrives at
\begin{equation}
p=\sum_{j=1}^{N}k_{j}=\sum_{j=1}^{N}2\pi \frac{m_{j}}{L}=2\pi \frac{n_{p}}{L}.
\label{total ang mom}
\end{equation}
$p$ is the eigenvalue of the total momentum operator, which
commutes with the Hamiltonian. Thus, the total momentum is quantized.
Since the expression for $p$ reduces to a sum over the $m_{j}$,
it is an invariant of the 'motion' as $c$ varies continuously from
$-\infty$ to $+\infty$. In particular, a state with zero angular
momentum for $c=0$ will always have zero angular momentum, independent of the value of $c$.

By inserting equation (\ref{Bethe ansatz}) into (\ref{exact energy}), one finds that the
full energy is given by the formula
\begin{equation}
E_L(c)=\sum_{j=1}^{N}k_{j}^{2}.\label{exact Energy}
\end{equation}
It contains also the energy of the interaction between particles and can therefore not be
considered as a purely kinetic energy.

There is one more important thing to note about the possible
$\{k_{j}\}$. If the set $\{ k_{j}\}$ is a solution, then the same is true for the
set $\{\tilde{k}_{j}\}$, defined by
\begin{equation}
\tilde{k_{j}}=k_{j}+2\pi n_{0}/L,
\end{equation}
which shifts the total angular momentum by $+2\pi n_{0} N/L$, where $n_0$ is an integer. 
Therefore, only states in the central momentum strip have to be considered, 
namely the states with
\begin{equation}
-\pi N/L<p\le \pi N/L.
\label{limitp}
\end{equation}

To obtain a solution, the set of equations (\ref{DEFINING EQS}) have to be solved together with (\ref{total ang mom}) and (\ref{limitp}).
Defining
\begin{equation}
\delta_{j}=(k_{j+1}-k_{j})L,\,\,\,\, j=1,...,N-1,
\label{AA}
\end{equation}
one changes variables from $\{k_{j}\}$ to $\{\delta_{j},p\}$.
Technically this can be done by introducing the vectors
\begin{equation}
{\bf k}=(k_1,...,k_N)^{\top},
\end{equation}
\begin{equation}
\boldsymbol{\delta}=(\delta_1,...,\delta_{N-1},pL)^{\top}
\end{equation}
and the $N$ by $N$ matrix
\begin{equation}
A=
\left( \begin{array}{cccccc}
-1 &  1 & 0  &     &     &       \\
 0 & -1 & 1  & 0   &     &       \\
   &  \ddots & \ddots & \ddots & \ddots   &       \\
   &    & 0  &  -1 &  1  &   0   \\
   &    &    &   0 & -1  &   1   \\
 1 & 1 & \dots& 1 &  1  &   1   
\end{array}
\right).
\label{A}
\end{equation}
The transformation is then given by substituting
\begin{equation}
{\bf k} = \frac{1}{L} A^{-1}\boldsymbol{\delta}
\label{subs}
\end{equation}
in (\ref{DEFINING EQS}).
Even when $N$ is large, the matrix $A$ can easily be inverted.
However, already for $N=4$ the set of equations (\ref{DEFINING EQS}) becomes so lengthy 
after this transformation that it makes no sense to give it here explicitly in terms of the new variables $\{\delta_j,p\}$.
Still, it can always be obtained with any computer algebra program.
In this work we used Mathematica \cite{Mathematica} and in all numerical calculations we used 
the new variables $\{\delta_{j},p\}$.

The order (\ref{order k_j}) implies that
$\delta_{j}\ge0$ when they are real.
It has been shown that in the limit $c\rightarrow 0$ also $c/\delta_j \rightarrow 0$ \cite{LiebLiniger}.
Hence, the arguments of all logarithms must approach unity in this limit and thus are far away 
from the branch cut discontinuity.
Therefore, if $|c|$ is small enough
\begin{equation}
\{n_1,...,n_{N-1}\}=\{n_1^0,...,n_{N-1}^0\}
\end{equation}
is certainly correct, where the $n_j^0$ denote the values of the $n_j$ for $c=0$. 
The $n_j^0$ still remain to be specified. 
This can be done by considering the non-interacting ground state.

\subsection{General Consequences drawn from the non-interacting ground state}
In the absence of interaction the ground state wave function is simply a constant.
This is equivalent to all $k_j$ and $p$ equal to zero.
Therefore, for $c=0$ one finds by
using (\ref{defTheta_jl})
\begin{equation}
e^{i\theta_{jl}}=-1,
\end{equation}
and with the choice of the ranges of $\theta_{jl}$ in (\ref{range theta_jl})
\begin{equation}
\theta_{jl}=-\pi,\,\,\,\,j>l,
\end{equation}
and
\begin{equation}
a(P)=1
\end{equation}
for any permutation $P$.
From (\ref{k_j}) it follows that
\begin{equation}
\delta_{j}=2\pi(m_{j+1}-m_{j}-1)=2\pi n_{j}^{0},\,\,\,\,j=1,...,N-1.
\label{delta_0}
\end{equation}
Hence all $n_j^0$ are
zero for the ground state. In the present work this is the only state that we are interested in.
The $n_j^0$ are unambiguously related to the $m_j$ via (\ref{delta_0})
and (\ref{total ang mom}). 
Therefore, they can be used equivalently to the $m_j$ to classify all states.
In this classification scheme the ground state is denoted by
\begin{equation}
\{n_1^0,...,n_{N-1}^0\}=\{0,...,0\}
\end{equation}
Thus, the non-interacting ground state determines the values of the integers $\{n_1^0,\dots,n_{N-1}^0\}$.

The numerical computation for the interacting ground state can now be started with 
the values $\{n_1,\dots,n_{N-1}\}=\{0,\dots,0\}$ for the quantum numbers $n_j$, provided $|c|$ is 
sufficiently small (see previous subsection).
However, the numerical effort can be reduced significantly by taking into account that all $n_j^0$ are equal for the ground state.
For $n_j^0=const$, it follows from the symmetry of the set of equations (\ref{DEFINING EQS})
that if $k$ is in $\{k_j\}$, then so is $-k$ \cite{LiebLiniger,MugaSnider}.
This is equally true for repulsive and attractive interactions.
Hence, for the ground state $p$ is equal to zero.
The order between the $k_j$ implies that $k_{N+1-j}=-k_j$ or equivalently $\delta_{N-j}=\delta_{j}$ for all $j$.
Hence, the number of variables is reduced for even $N$ to
$N/2$ and for odd $N$ to $(N-1)/2$.
This is also the number of equations that remain to be solved,
as can be seen by substituting these equalities into (\ref{DEFINING EQS}), and therefore we refer to this number as $N_{eqs}$.
Therefore, the problem consists now of solving $N_{eqs}$ coupled transcendental equations in $N_{eqs}$ unknowns.
The numerical solution of these equations is discussed in section \ref{NumericalSolution}.

It is important to note that only the $n_j^0$ and not the $n_j$ may be taken for a unique classification of states
since the latter may change when $c$ is decreased or increased from zero onwards.
In fact, for $c>0$ one finds numerically that the arguments of all logarithms in (\ref{DEFINING EQS})
move clockwise around zero on the unit circle when $c$ increases.
Hence, each of the $n_j$ has to be changed to $n_j+1$ 
when the argument of the $j$-th logarithm crosses the branch cut discontinuity of the principal part of the 
logarithm.
For $c<0$, it turns out that the ground state $k_j$ are purely imaginary and
hence the arguments of all logarithms in (\ref{DEFINING EQS}) remain positively real.
This implies that for attractive interaction the $n_j$ always 
remain at their values $n_j^0$ for zero interaction strength.
Since the $k_j$ are purely imaginary for attractive interaction, it is customary to use the variables $\{\alpha_j,p\}$ for $c<0$,
where
\begin{equation}
\alpha_j=i\delta_j,\,\,\,\, j=1,\dots,N-1.
\label{Alpha_j}
\end{equation}

\subsection{Perturbation theory}
For sufficiently weak repulsive or attractive interaction perturbation theory should also be applicable.
Since the non-interacting ground state is just a constant, it is easy to derive the energy to first order.
By treating the whole interaction potential
\begin{equation}
2c\sum_{i<j}^N \delta(x_i-x_j)
\end{equation}
as a small perturbation of the unperturbed ground state with the wave function
\begin{equation}
\psi(x_1,...,x_N)=\left(\frac{1}{\sqrt{L}}\right)^N,
\end{equation}
one finds by using non-degenerate first order perturbation theory that the ground state energy per particle is given by
\begin{equation}
\frac{E^{(1)}_{L}(c)}{N}=\frac{c(N-1)}{L}.
\label{1storderpert}
\end{equation}

\section{Numerical Solution \label{NumericalSolution}}
In this section we discuss the numerical solution of the set of coupled equations (\ref{DEFINING EQS}) starting
from the non-interacting ground state.
By solving (\ref{DEFINING EQS}),
it is assumed that the ground state takes on the form of a
\emph{Bethe-ansatz} wave function.
For repulsive interaction the validity of the \emph{Bethe-ansatz} has been proven rigorously \cite{YangYang,Dorlas}.
However, for attractive interaction there is no such proof to our knowledge.
As discussed in section \ref{Theory}, for two and three bosons there are known cases \cite{LiebLiniger,MugaSnider}
in which the form of the wave function is \emph{not} of the \emph{Bethe-ansatz} type for certain critical, 
attractive interaction strengths.
However, for the ground state this assumption is justified further below 
by using an argument similar to that of Muga and Snider \cite{MugaSnider}.
For all computations we used the Mathematica function \emph{FindRoot} \cite{Mathematica} which makes 
use of the Newton-Raphson algorithm.
All calculations were carried out on a standard PC.

We define the residual error $\Delta_{res}$ as the sum over the absolute values of
the differences between the left- and the right-hand sides of each of the equations
(\ref{DEFINING EQS})
\begin{equation}
\Delta_{res}=\sum_{j=1}^{N_{eqs}} |\Delta_{j}|,
\end{equation}
where $\Delta_j$ is the difference in the $j$-th equation.
Our goal is a residual error of
\begin{equation}
\Delta_{res} \le 10^{-9}
\end{equation}
in all computations. Thus, we ensure that the wave vectors $k_j$ are accurate up 
to the eighth digit. We found that for repulsive interaction this goal is fairly easy to achieve,
whereas the attractive case proved to be a lot more problematic.
In the numerical calculation, we started from the non-interacting ground state,
and increased/decreased $c$ stepwise. For very weak interaction and small particle numbers 
the $FindRoot$ function is not too sensitive to the initial guess.
As an initial guess for the following sets of $\{\delta_j\}$ or $\{\alpha_j\}$
we used the three previously calculated sets and extrapolated to the next one.

\begin{figure}
\includegraphics[width=11cm, angle=-90]{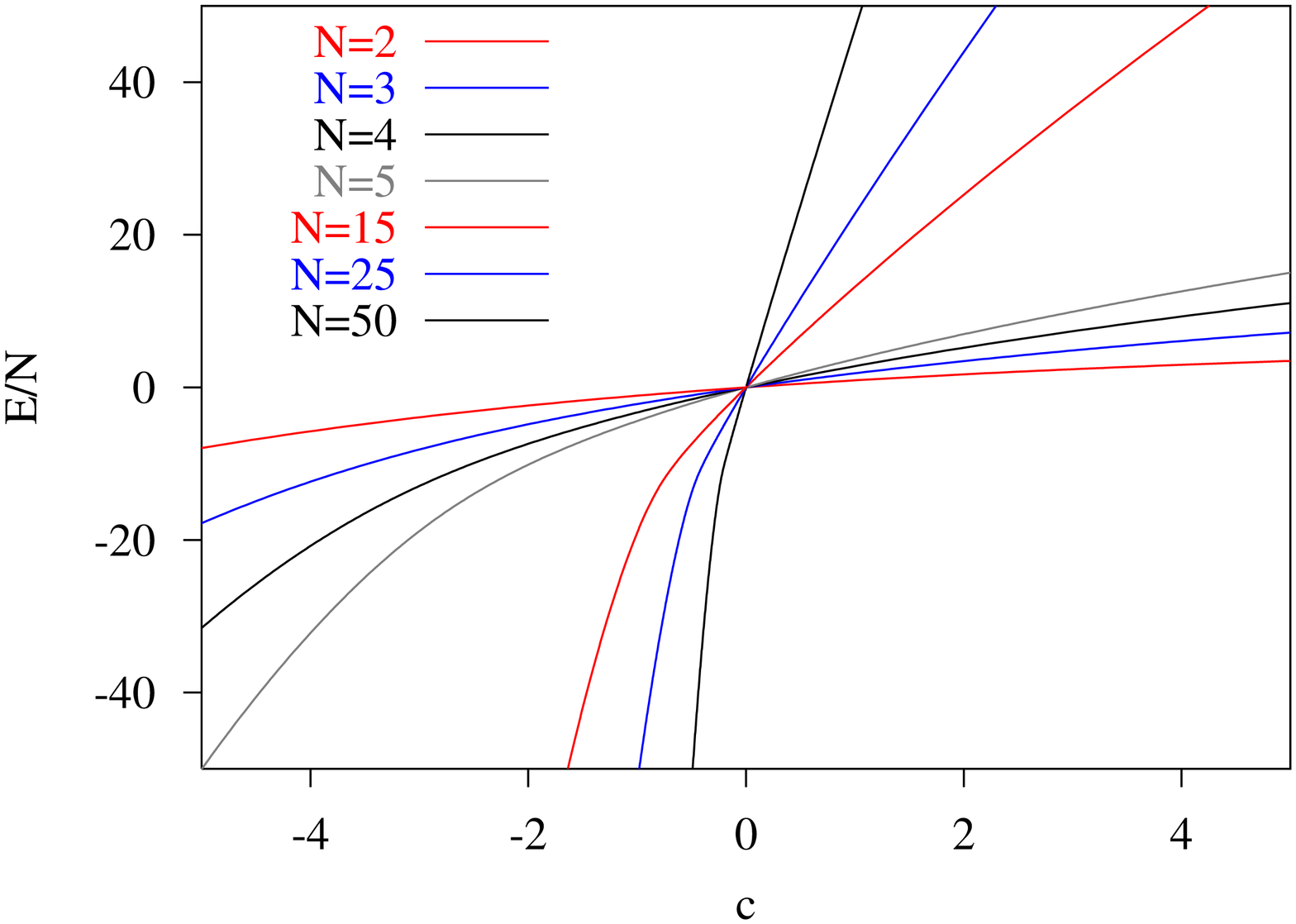}
\caption{(color online) Exact ground state energies per particle as a function of the interaction strength $c$.
The \emph{Bethe-ansatz} was used to compute the solutions to the problem of $N$ bosons on a ring of length $L$ for 
repulsive $c>0$ and attractive $c<0$ interaction strengths.
It can be seen that the ground state energies for repulsive interaction are smoothly connected at $c=0$ to 
those of the \emph{Bethe-ansatz} solutions for attractive interaction with the same particle and quantum numbers (see text). 
This proves the validity of the \emph{Bethe-ansatz} also for attractive interaction for all cases studied.
The ring was taken to be of unit length $L=1$.}
\label{fig1}
\end{figure}

\begin{figure}
\begin{center}
\centering
\subfigure{\epsfig{figure=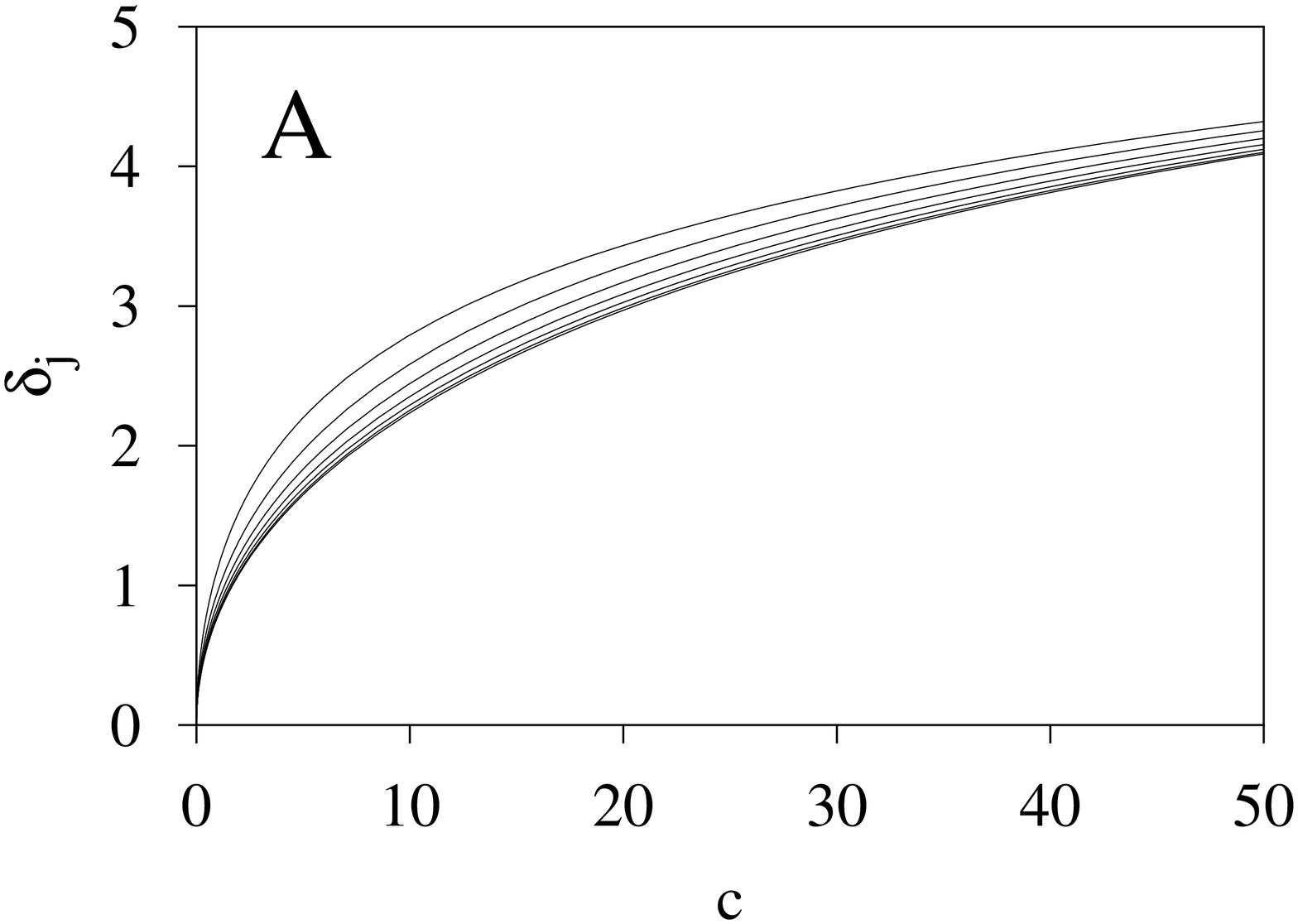,width=7cm, angle=-90}} \\
\subfigure{\epsfig{figure=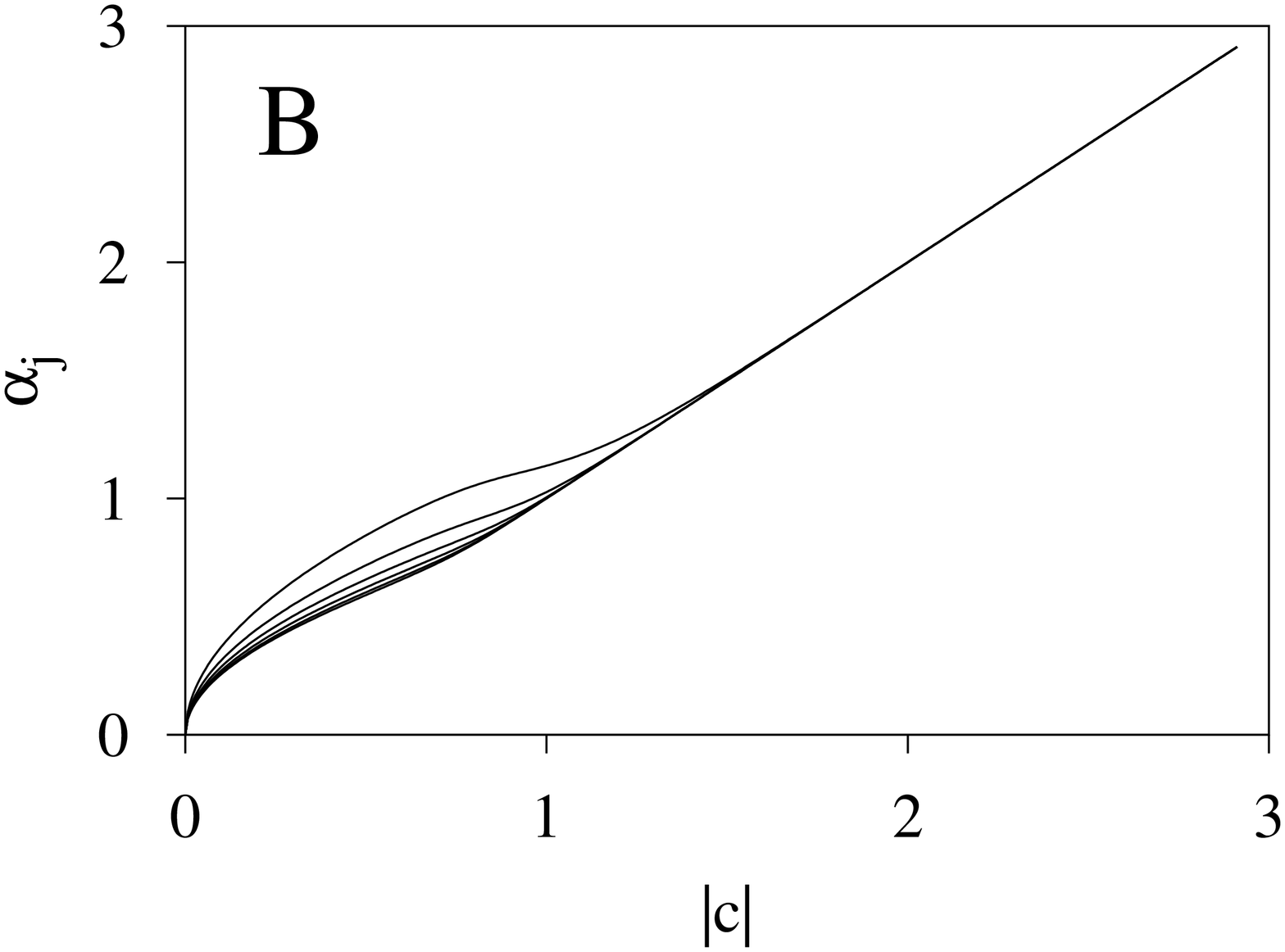,width=7cm, angle=-90}}
\end{center}
\caption{Lieb and Liniger's set of coupled transcendental equations is solved by the set of variables $\{\delta_j\}$. 
For the repulsive ground state all $\delta_j$ are real, whereas they are purely imaginary for the attractive one.
This motivates the use of the variables $\alpha_j=i\delta_j$ for attractive interaction.
Depicted are the sets $\{\delta_j\}$ and $\{\alpha_j\}$ for fifteen bosons on a ring of unit length.
In the absence of interactions all $\delta_j$ are zero.
A: Repulsive interaction.  Starting from zero the $\delta_j$ begin to spread and
are well separated from one another
before approaching their limiting value of $2\pi$.
B: Attractive interaction. The $\alpha_j$ spread, 
but start to degenerate already for comparatively 
weak interaction. This complicates the numerical solution significantly.} 
\label{fig2}
\end{figure}

The computed ground state energies per particle are depicted in Fig. \ref{fig1} as a function 
of the interaction strength $c$.
It is seen that the energies of the 
repulsive ground state solutions are \emph{smoothly} connected at $c=0$ to the energies per 
particle of the corresponding \emph{Bethe-ansatz} solutions for attractive interaction. 
The smooth connection of the energies is by no means trivial, as we shall show now.
As mentioned in section \ref{Theory}, the $\delta_j$ are purely real 
for repulsive interaction, but purely \emph{imaginary} for attractive, motivating the 
redefinition of variables (\ref{Alpha_j}). This can be seen for the case of fifteen particles in Fig. \ref{fig2}, 
which we shall discuss representatively for all other particle numbers.
Since the $\delta_j$ are real for $c>0$ and imaginary for $c<0$ they can not be smoothly connected at $c=0$.
It is therefore quite surprising that the energy, which is a function of these variables, is smoothly connected at $c=0$.

For repulsive and attractive interactions it can be seen that all $\delta_j$ and all $\alpha_j$ 
respectively converge to zero when $|c|\rightarrow 0$. Together with $p=0$ this implies that the 
non-interacting ground state for which all $k_j$ are zero is approached from either side.

In the repulsive case all $\delta_j$ start from zero and begin to spread when $c$ 
is increased from zero onwards. After a maximal spread the $\delta_j$ start to 
degenerate successively and saturate to their value at infinity, 
namely $2\pi$ \cite{LiebLiniger}.

For attractive interaction the variables $\alpha_j$ are used.
The $\alpha_j$ behave similar to the repulsive case in the 
sense that they first start from zero,  spread and then degenerate,
but there is no saturation for strong attractive interaction.
These findings are equally true for all particle numbers studied in this work.

The smooth connection of the energies and the continuous connection of the $k_j$ at $c=0$
prove the validity of the \emph{Bethe-ansatz} numerically for the ground state of up to fifty bosons. 
However, we suspect that a complete set of states can be obtained for any particle number
from a $\emph{Bethe-ansatz}$ wave function 
in the sense explained in section \ref{Theory}.

Numerically, the situation is very different for repulsive interaction and attractive interaction. 
The degeneracy sets in much earlier for attractive than for repulsive interaction. 
This is important for the following reason.
We found that the numerical calculation tends to break down,
if the relations
\begin{equation}
\delta_1>\delta_2>...>\delta_{N_{eqs}},
\label{order delta_j}
\end{equation}
\begin{equation}
\alpha_1>\alpha_2>...>\alpha_{N_{eqs}}
\label{order alpha_j}
\end{equation}
for repulsive and attractive interaction, respectively, are not fulfilled at all times.
This illustrates the importance of a proper initial guess.
The graphs of the $\{\delta_j\}$ and $\{\alpha_j\}$ for other particle numbers 
are very similar to Fig. \ref{fig2}. We discuss the effects that arise due to changing $N$ 
in section \ref{resultsonrepulsivebosons} and section \ref{resultsonattractivebosons}.

Due to the degeneracy of the $\delta_j$ and the $\alpha_j$ we were forced to perform all calculations with high precision numbers.
The Mathematica option \emph{WorkingPrecision} \cite{Mathematica} allows for computations with numbers of arbitrary precision. 
For repulsive interaction we found that it is sufficient to work with a number precision of no more than $10^{-20}$ for all variables 
involved in the computation, at least for up to fifty particles and $c<1500$. 
Then, the smallest difference between the $\delta_j$, $\delta_{24}-\delta_{25}$, 
is still greater than $10^{-6}$.
For attractive interaction on the other hand we found that we had to use numbers of extremely high precision
in the calculations for fifty particles, even for $|c|<1$. 
Then, the difference $\alpha_{24}-\alpha_{25}$ is as little as $10^{-85}$.
For fifty particles we could achieve our goal of a residual error of less than $10^{-9}$ only by using 
numbers with $90$ digits!
However, the fact that all computations were carried out on a standard PC proves
that the possibilities of our approach are still far from exhausted.

\section{Results on repulsive bosons \label{resultsonrepulsivebosons}}
\subsection{Exact finite $N$ solution for repulsive interaction}
\begin{figure}
\includegraphics[width=11cm, angle=-90]{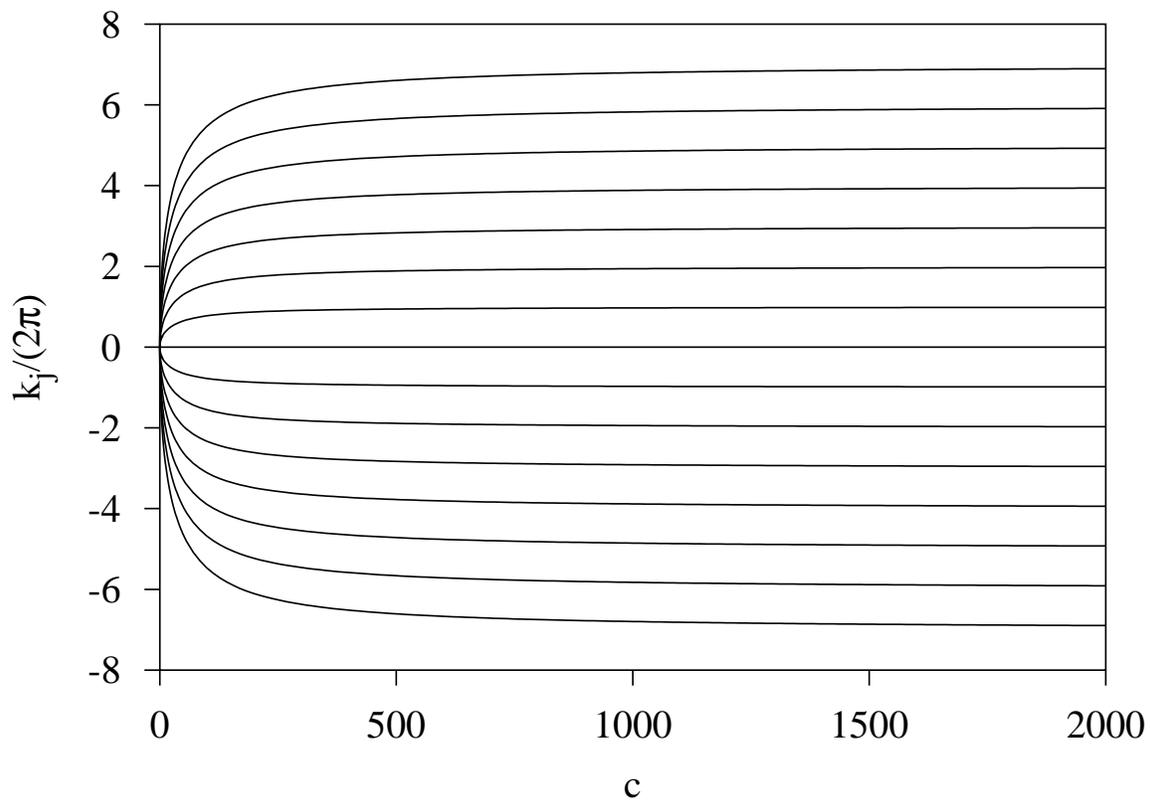}
\caption{Wave vectors $k_j$ for fifteen repulsive bosons on a ring of unit length.
Only for very strong interaction the wave vectors are practically equally spaced with a separation 
of $2\pi$ between adjacent $k_j$, also see Fig. \ref{fig2}A.}
\label{fig3}
\end{figure}

By solving the set of coupled equations (\ref{DEFINING EQS}) for $c>0$, as described in section \ref{NumericalSolution}, we find for each
$c$ the corresponding set $\{\delta_j,p\}$. Using (\ref{subs}), one obtains the corresponding set $\{k_j\}$.
In Fig. \ref{fig3} the set $\{k_j\}$ is depicted for the case of fifteen bosons. 
Starting from the non-interacting ground state, where the wave function 
is a constant and all $k_j$ are zero, the $k_j$ spread for $c>0$. 
In the limit of very strong interaction we recover Lieb and Liniger's result 
that all $k_j$ become constant and equally spaced with a separation of $2\pi$ between adjacent
$k_j$. 
The expression for the energy (\ref{exact Energy}) then implies that the energy per particle 
saturates to a finite value when $c\rightarrow\infty$.
For other particle numbers the situation is much the same, only that the larger the number of particles, 
the later this saturation sets in. 

\subsection{The relation to the Tonks-Girardeau limit}
\begin{figure}
\begin{center}
\subfigure{\epsfig{figure=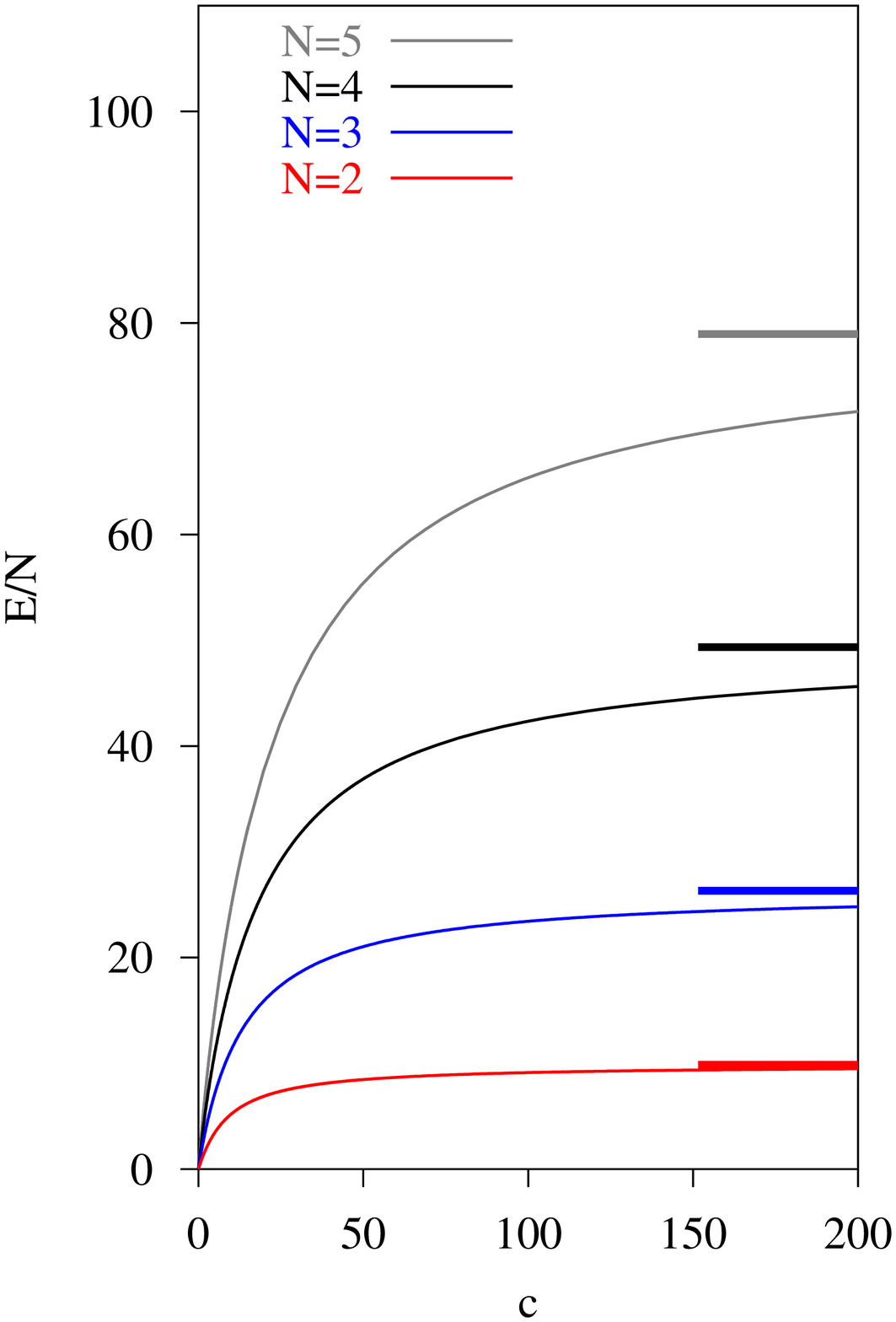,width=7cm, angle=0}} 
\subfigure{\epsfig{figure=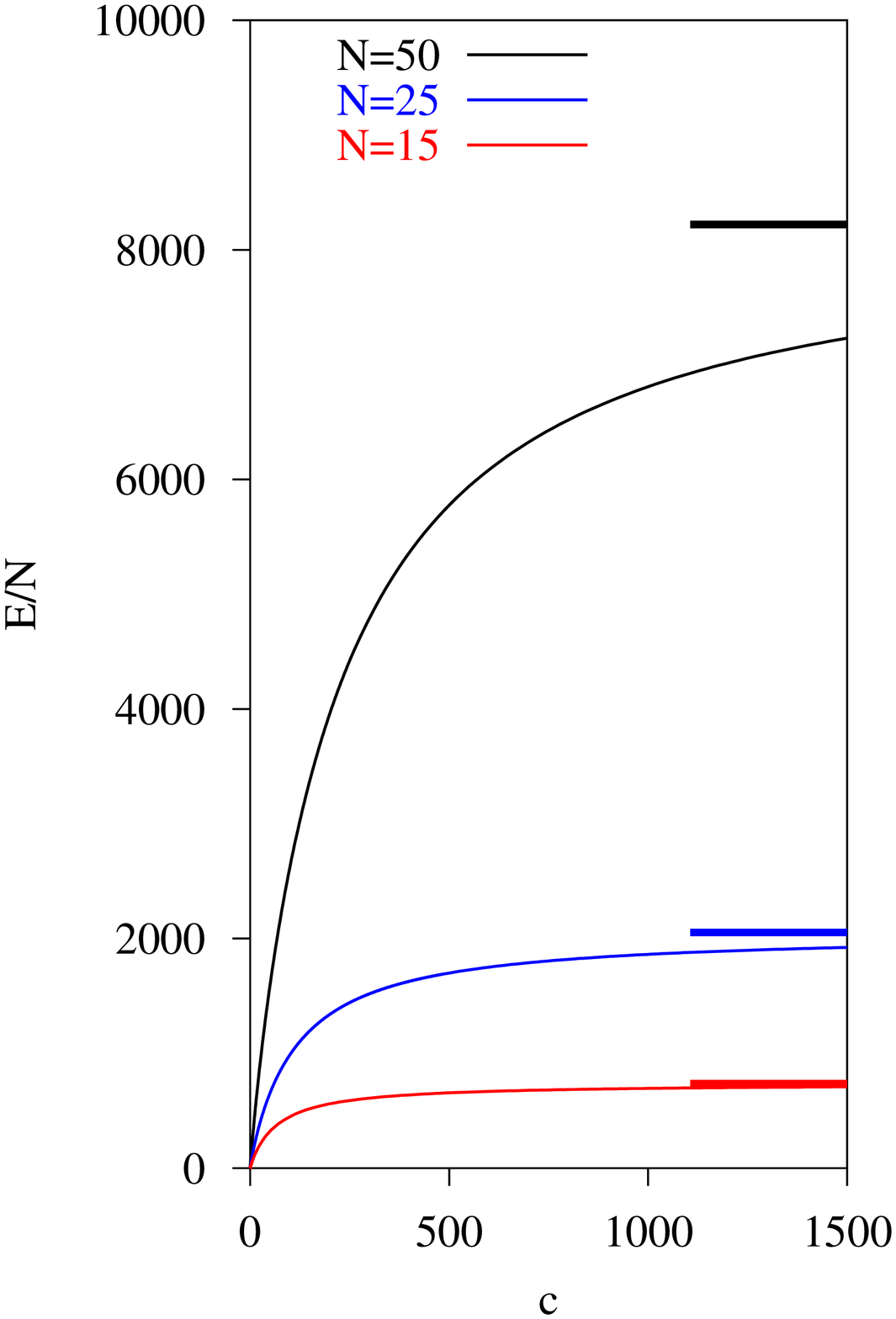,width=7cm, angle=0}}
\end{center}
\caption{(color online) Ground state energies per particle for different $N$ on a ring of \emph{constant} length ($L=1$) for repulsive interaction ($c>0)$.
For $c\rightarrow \infty$ the energy per particle converges to that of the Tonks-Girardeau (TG) limit. 
On the right border of each graph the energy of the TG limit is indicated for each particle number.
For larger particle numbers the TG limit is approached for larger values of $c$.
%If the relative difference of the TG energy to the exact energy (see definition in the text and equation (\ref{frac})) 
%is to remain constant when the number of particles is increased, 
%also $c$ has to be increased. 
%See also Fig. \ref{fig5}.
}
\label{fig4}
\end{figure}

The saturation for strong interaction can be quantified by considering 
the energy difference to the saturation energy. 
From a physical argument it is clear that the saturation energy has to coincide 
with the energy per particle of the TG gas which is given by \cite{Girardeau}:
\begin{equation}
\frac{E_{L,TG}}{N}=\frac{(N^2-1)\pi^2}{3 L^2},
\label{Girardeau}
\end{equation}
where $L$ is the length of the ring.
In Fig. \ref{fig4} the exact energies per particle on a ring of unit length 
are plotted as a function of $c$ and the TG energy is 
indicated on the right border of the graph for different particle numbers. 
It can be seen clearly that the larger the number of particles is, the stronger the interaction has to be 
to reach a given fraction of the TG energy.

We would like to answer the following question quantitatively: What is the smallest $c$ for which 
the TG limit is less than a certain percentage $r$ away from the exact energy
of the system?
This relative deviation is given by
\begin{equation}
\left(\frac{\Delta E}{E}\right)_{L,TG}(c)=\frac{E_{L,TG}-E_{L}(c)}{E_{L}(c)}
\label{frac}
\end{equation}
and thus we are looking for the $c$ values $c_r$, for which
\begin{equation}
\left(\frac{\Delta E}{E}\right)_{L,TG}(c_{r})=r.
\label{equr}
\end{equation}
This relative deviation may also be considered as the error in the energy introduced 
by approximating the $N$ bosons on a ring at finite $c$ by the TG limit.

\begin{figure}
\includegraphics[width=11.2cm, angle=0]{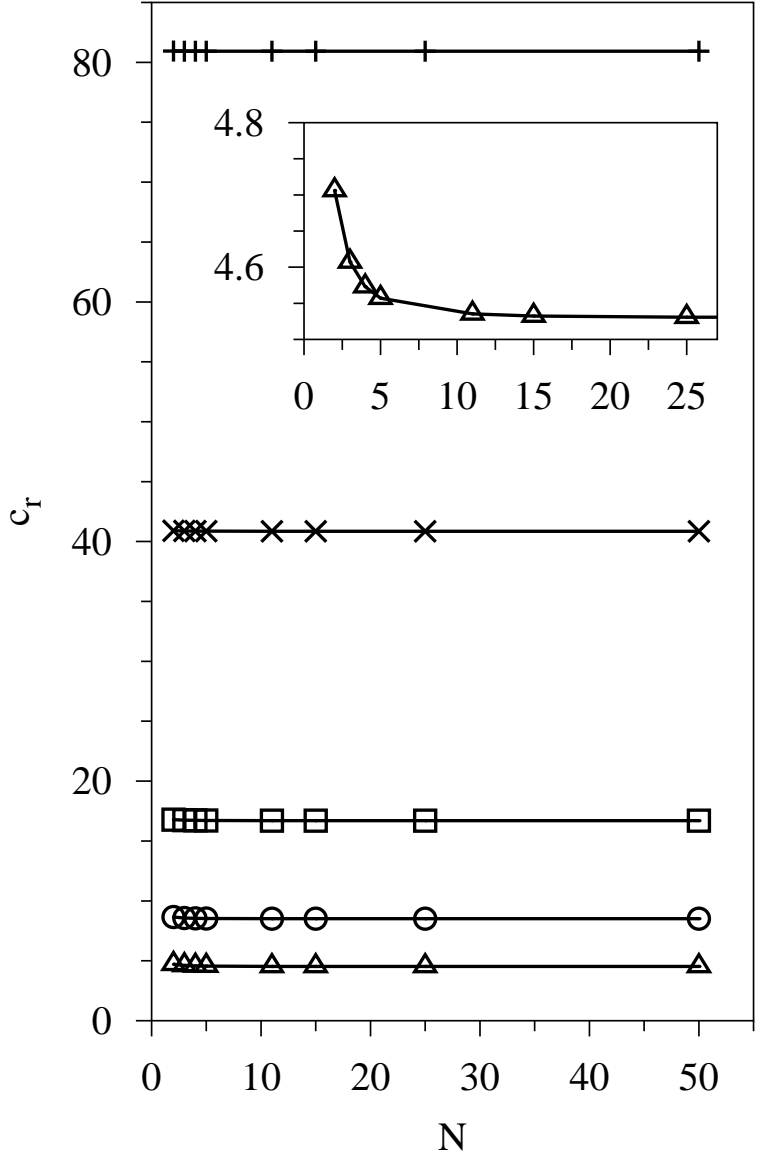}
\caption{Measure for the TG energy to approximate the exact energy.
Given $N$ bosons on a ring of \emph{variable} length,
at the interaction strength $c=c_r$ the relative difference of the TG energy to the exact energy equals the fraction $r$
(see definition in the text and equation (\ref{frac})).
For this graph the density $\rho=N/L$ was held constant, $\rho=1$. 
Surprisingly, for a given $r$ the values $c_r$ are practically 
independent of the number of particles when $\rho$ is a constant. 
Even more surprising is the fact that this is equally true for small and large values of $r$. 
Top to bottom: $r=5\%, 10\%, 25\%, 50\%, 95\%$. The inset shows how little the deviation from a constant is, even for 
$r$ as large as  $95\%$.
The corresponding $c_r$ for a ring of length $L=1$ can be obtained
by multiplying each $c_r$ value by $N$. }
\label{fig5}
\end{figure}

First, we discuss the scaling properties of $\left(\frac{\Delta E}{E}\right)_{L,TG}(c)$  with respect to $L$. 
By using (\ref{scaleEnergy}) and (\ref{Girardeau}) one finds that
\begin{equation}
\left(\frac{\Delta E}{E}\right)_{L^{\prime},TG}(c)= \left(\frac{\Delta E}{E}\right)_{L,TG}(c\frac{L^{\prime}}{L}).
\label{Gircale}
\end{equation}
Relation (\ref{Gircale}) allows to calculate the values $c_r$ for a ring of length $L^{\prime}$, if $c_r$ is
known for a ring of length $L$. The result is shown in Fig. \ref{fig5}.
For illustration purposes we plot the $c_r$ values for a constant density $\rho=N/L=1$, rather than a constant length of the ring.
To get the corresponding $c_r$ values for a ring of length $L=1$,
one simply needs to multiply each $c_r$ value by $N$.
Fig. \ref{fig5} shows that the $c_r$ are almost \emph{independent} of $N$ 
for \emph{any} given $r$, when the density is constant.
For a ring of length $L$, this implies that 
the corresponding $c_r$ depend almost linearly on the number of particles.
The slopes of the curves for the $c_r$ on a ring of unit 
length as a function of $N$ are exactly the $c_r$ in Fig. \ref{fig5}.
The inset shows how little the deviation 
from a constant slope is, even for $r$ as large as 95\%. 
When $r$ is decreased this deviation becomes even smaller.
The fact that the TG limit is reached for larger $c$-values with increasing $N$ can therefore solely be attributed
to the increase in the density by adding more particles.
Using (\ref{Girardeau}) and Fig. \ref{fig5} one 
can even obtain a rough estimate to the 
exact ground state energy for \emph{any} particle number, simply 
by extrapolating for the desired $N$ on 
the graph to $c$, multiplying $c$ times $N$ and solving (\ref{equr}) for $E_{L=1}(c)$.
It is surprising, how simple the relation between the system for finite $c$ and the TG limit is at constant density. 
There is  almost no dependence of the $c_r$ on the particle number, especially when $r$ is small and $N$ is more than just a 
few particles.

\subsection{The relation to the thermodynamic limit}
Now we address the question, how far away the thermodynamic limit \cite{LiebLiniger} is from the 
finite $N$ solution.
In the thermodynamic limit $N,L\rightarrow\infty$ with $\rho=N/L=const$, 
and the energy of this solution differs from the exact energy
of the finite $N$ system, even if the densities are the same.
The energies $E_{\rho=1,TDL}/N$ of this solution can be found elsewhere \cite{Olshaniiwebdata}.
For large particle numbers, $N\gg1$, it is expected that the thermodynamic limit energy 
gives a good approximation to that of the finite $N$ system, provided the densities are the same.
For small $N$, however, the finite number of particles should 
play an important role.
We would like to answer the following questions.
Firstly, how large does $N$ have to be in order that the thermodynamic limit
solution provides a good approximation to the finite system, and secondly,
what is the quality of this thermodynamic limit approximation, 
when $c$ varies for a given $N$?

\begin{figure}
\begin{center}
\centering
\subfigure{\epsfig{figure=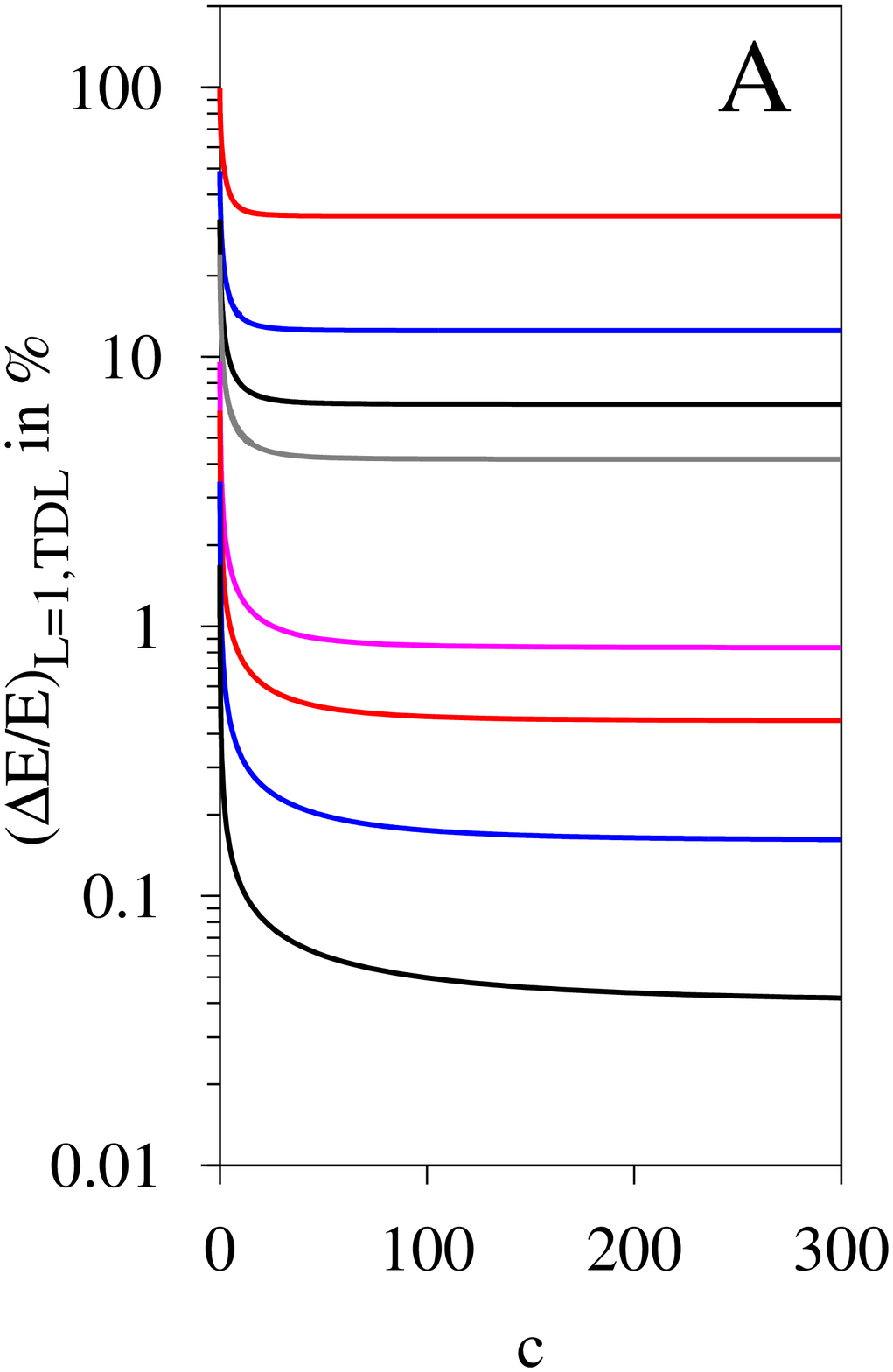,width=7cm, angle=0}} 
\subfigure{\epsfig{figure=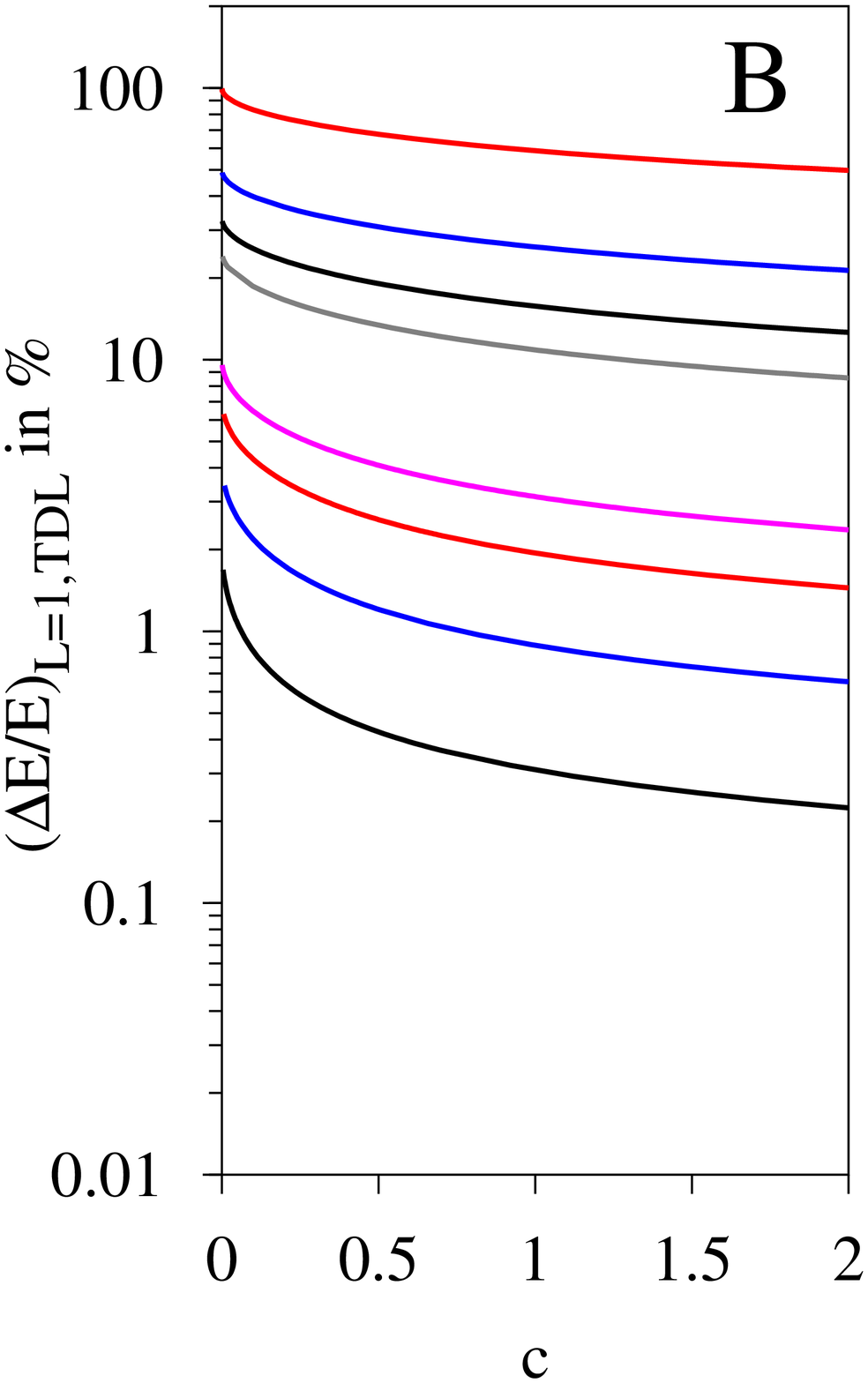,width=7cm, angle=0}}
\end{center}
\caption{(color online) Measure for the thermodynamic limit to approximate the exact solution.
Lieb and Liniger's solution in the thermodynamic limit is often used as an approximation to the true finite $N$ solution.
This introduces an error which depends on the number of particles, 
the interaction strength and the length of the ring. In the limits $c\rightarrow 0$ 
and $c\rightarrow \infty$ the error in the energy can be calculated by using perturbation theory (see text) and the TG expression for the 
energy (see text). 
The formulas obtained for infinite and zero interaction strength, see Eqs. (\ref{TDLerrorcinfty}) and 
(\ref{TDLerrorczero}) respectively, imply that this error is \emph{never} less than $1\%$ if $N<11$.
Similarly, only for $N>101$ it is \emph{always} less than $1\%$. These results do not depend on the length of the ring, (see text).
Curves from top to bottom: $N=2,3,4,5,11,15,25,50$. Length of the ring: $L=1$.
A: The error introduced by using the thermodynamic limit solution instead of the solution for finite $N$ is a monotonously decreasing 
function of the interaction strength and the number of particles. In the limit $c\rightarrow \infty$ the error approaches 
the value $\frac{1}{N^2-1}$.
B: For $c\rightarrow 0$ the error converges to the finite value $\frac{1}{N-1}$ in agreement with perturbation theory.
} 
\label{fig6}
\end{figure}

To answer these questions quantitatively
we consider the energy of the thermodynamic limit solution for the density
$\rho=N/L=N$ and compare with our results on a ring of unit length. 
This ensures that we compare the two systems at the same density.
We define the relative deviation of the thermodynamic limit energy from the energy of the finite $N$ solution 
\begin{equation}
\left(\frac{\Delta E}{E}\right)_{L=1,TDL}(c)=\frac{E_{\rho=N,TDL}(c)-E_{L=1}(c)}{E_{L=1}(c)}.
\label{TDLerror}
\end{equation}
This quantity can also be considered as the error which is introduced by using the energy
of the thermodynamic limit solution instead of that of the  
finite $N$ solution. 
Again, we first discuss the scaling properties of this relative deviation when $L$ varies.
The ground state energy in the thermodynamic limit can be written as \cite{LiebLiniger}:
\begin{equation}
E_{\rho,TDL}(c)=N \rho^2 e(\frac{c}{\rho}),
\end{equation}
where $e(x)$ is a monotonically increasing function
which is tabulated in \cite{Olshaniiwebdata}.
Therefore, we find the same scaling behaviour as for the TG 
limit (\ref{Gircale}), as follows by using (\ref{scaleEnergy}) 
\begin{equation}
\left(\frac{\Delta E}{E}\right)_{L^{\prime},TDL}(c)=\left(\frac{\Delta E}{E}\right)_{L,TDL}(c\frac{L^{\prime}}{L}).
\label{scaleTDLerror}
\end{equation}
The relative deviation $\left(\frac{\Delta E}{E}\right)_{L,TDL}(c)$ is shown in Fig. \ref{fig6} for $L=1$.
At first sight, the result is quite surprising.
For all particle numbers the relative deviation is \emph{largest} for $c=0$, but decreases rapidly before
saturating to a \emph{finite} value for infinitely strong interaction.
However, one can explain this behaviour by considering
the limiting cases $c\rightarrow \infty$ and
$c\rightarrow 0$.

We begin with the limit $c \rightarrow \infty$.
In the thermodynamic limit the impenetrable boson gas energy 
per particle is given by \cite{Girardeau}:
\begin{equation}
\frac{E_{\rho,c=\infty}}{N}=\frac{\pi^2}{3}\rho^2,
\label{GirardeauTDL}
\end{equation}
whereas (\ref{Girardeau}) is the $c\rightarrow\infty$ limit of the finite $N$ system.
Substituting $L=1$ in (\ref{Girardeau}) and $\rho=N$ in (\ref{GirardeauTDL}) results in
\begin{equation}
\left(\frac{\Delta E}{E}\right)_{L=1,TDL}\stackrel{c\rightarrow\infty}{\longrightarrow}\frac{1}{N^2-1},
\label{TDLerrorcinfty}
\end{equation}
which is exactly the tendency that can be seen in Fig. \ref{fig6}A.

For zero interaction strength the behaviour can be explained
by treating the full interaction between particles 
as a small perturbation to the non-interacting ground state for finite $N$, 
which results in (\ref{1storderpert}) with $L=1$.
The corresponding leading term of the energy per particle in the thermodynamic 
limit for small $c$ is given by \cite{LiebLiniger}:
\begin{equation}
\frac{E_{\rho=N,TDL}(c)}{N}=cN.
\end{equation}
Hence, the relative deviation $\left(\frac{\Delta E}{E}\right)_{L=1,TDL}$ for $c\rightarrow 0$ becomes 
\begin{equation}
\left(\frac{\Delta E}{E}\right)_{L=1,TDL}\stackrel{c\,\rightarrow 0}{\longrightarrow}\frac{1}{N-1}.
\label{TDLerrorczero}
\end{equation}
as can be seen in Fig. \ref{fig6}B.

An interesting consequence of (\ref{scaleTDLerror}) is that for given $N$ and $L$,
the relative deviation $\left(\frac{\Delta E}{E}\right)_{L,TDL}(c)$ is obtained
by the value of the curve for the same $N$ in Fig. \ref{fig6},
evaluated at $\tilde{c}=Lc$.
Varying the length of the ring only changes the point of evaluation in Fig. \ref{fig6}.
Consequently, neither (\ref{TDLerrorcinfty}) nor (\ref{TDLerrorczero}) depend on the length of the ring.
Obviously, for small particle numbers the thermodynamic limit approximation is 
never a good approximation, no matter how strong the 
interaction or how large the size of the ring is. 

\subsection{Quality of the thermodynamic limit approximation}
\begin{figure}
\includegraphics[width=11cm, angle=-90]{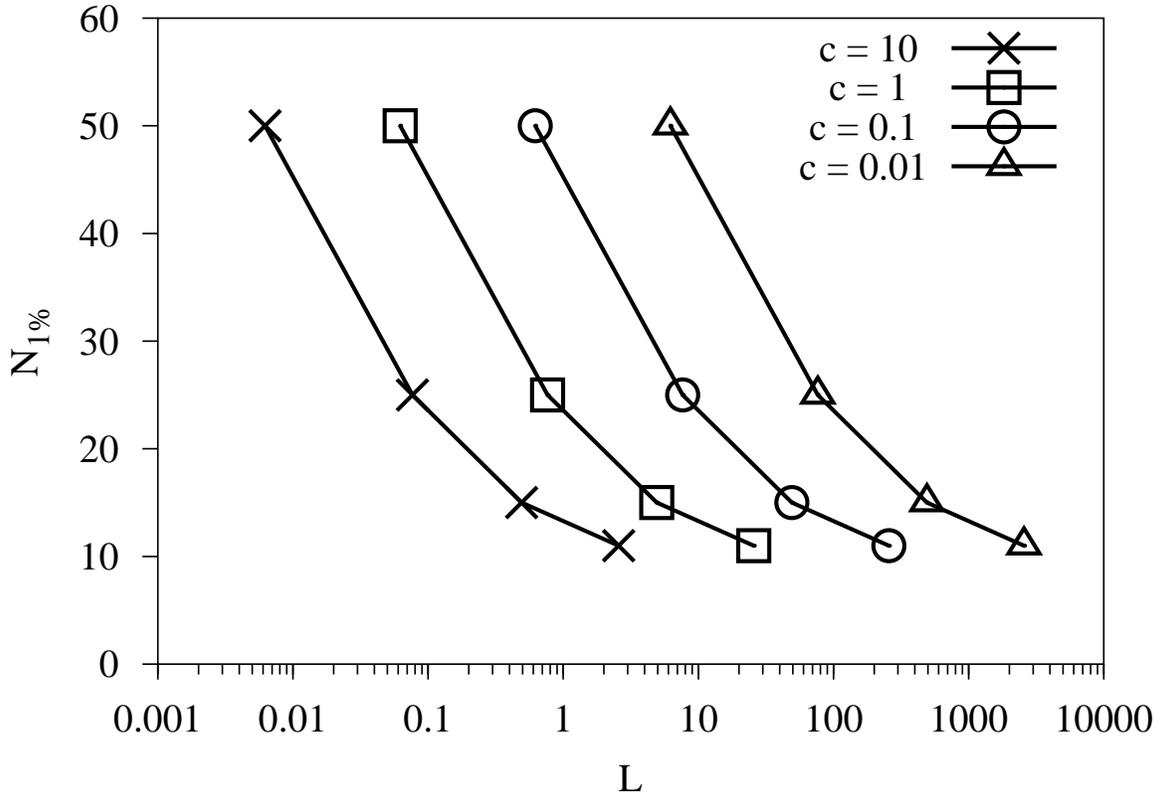}
\caption{Depicted are those $N$ for which the error in the energy introduced by using the thermodynamic limit solution instead of the finite $N$ solution is less than one percent. These $N$ are denoted by $N_{1\%}$.
Equation (\ref{scaleTDLerror}) implies that the curves can be shifted horizontally, in the sense that a simultaneous change of $L\rightarrow x L$ 
and $c\rightarrow \frac{1}{x} c$ does not change $N_{1\%}$. It can be seen that there is no $N_{1\%}<11$ since for $N<11$ the error of the 
thermodynamic limit approximation never drops below one percent (see also Fig. \ref{fig6}).
}
\label{fig7}
\end{figure}

In the context above, the following question arises naturally.
Given a certain interaction strength $c$ and a ring of a fixed size $L$,
how many particles are \emph{at least} necessary for the thermodynamic 
limit approximation to be accurate to a certain percentage, for instance 1\%?
We denote this number by $N_{1\%}$. The result is shown Fig. \ref{fig7}.
For weak interaction strength the number of particles has to be much larger than for 
strong interaction. Although this is counterintuitive, one should remember that weak 
interaction is the regime where the thermodynamic limit approximation is worst, see Fig. \ref{fig6}.
Equivalently, for a given particle number, the size of the ring has to be much larger
for weak interaction strength than for strong interaction, if the thermodynamic limit approximation is 
to be used for the description of the experiment. 
It is no coincidence that the curves in Fig. \ref{fig7} all look the same.
From (\ref{scaleTDLerror}) it follows that if one changes $L \rightarrow L^{\prime}$ and simultaneously
$c\rightarrow \frac{L}{L^{\prime}} c$, then the deviation $\left(\frac{\Delta E}{E}\right)_{L,TDL}$ remains unchanged.
Therefore, the curve, e.g., for $c=0.1$ can be obtained from the curve for $c=1$ by multiplying
all $L$ values times ten. This permits to obtain curves for all values of $c$.

From (\ref{TDLerrorcinfty}) it follows that for $N\le10$, 
the relative deviation $\left(\frac{\Delta E}{E}\right)_{L=1,TDL}$ does not drop below
one percent for any $c$. Moreover, from (\ref{scaleTDLerror}) it follows that this can not be 
compensated for by making the ring larger. We conclude that the thermodynamic limit 
approximation can never be accurate to $1\%$, if the number of particles is less than $11$.
Similarly, it follows from (\ref{TDLerrorczero}) that only for $N>101$ the relative deviation $\left(\frac{\Delta E}{E}\right)_{L,TDL}$ 
is always less than $1\%$.

\section{Results on attractive bosons \label{resultsonattractivebosons}}
\subsection{Exact finite $N$ solution for attractive interaction}
\begin{figure}
\includegraphics[width=11cm, angle=-90]{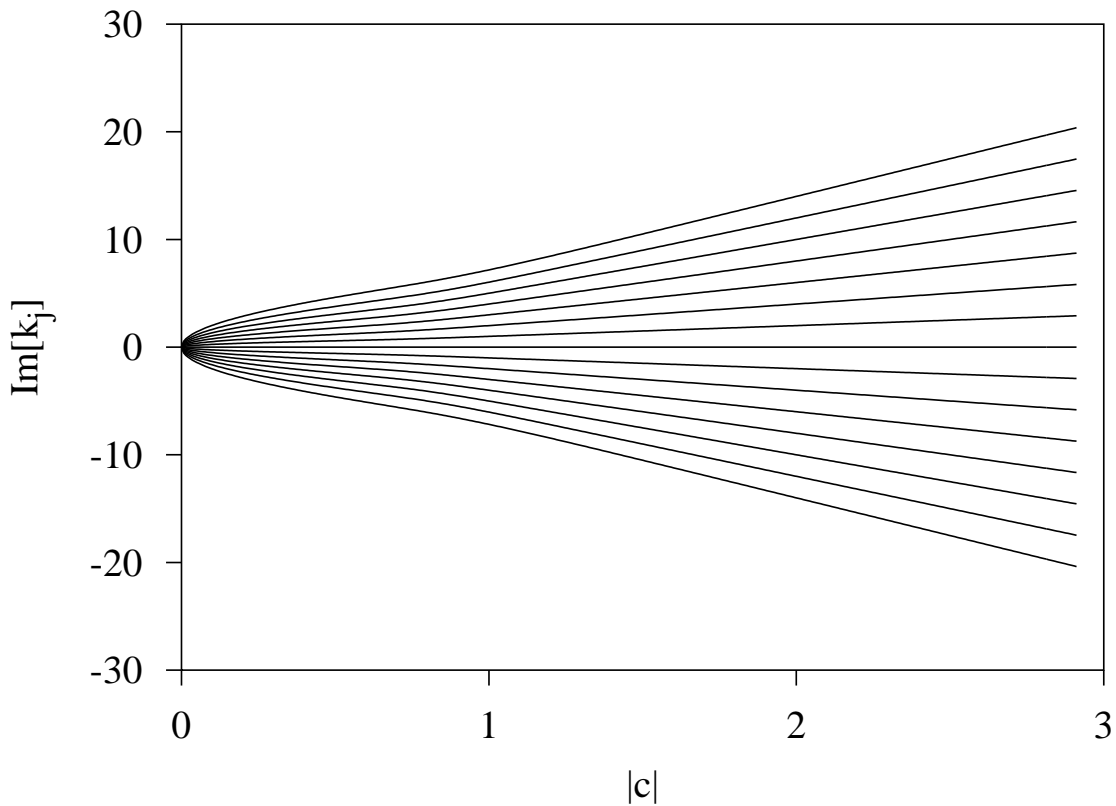}
\caption{The ground state wave vectors $k_j$ of fifteen bosons
for attractive interaction on a ring of unit length. All $k_j$ are purely imaginary.
In contrast to the repulsive case, see Fig. \ref{fig3}, there is no saturation for strong attractive interaction. 
For strong attractive interaction all $k_j$ grow (approximately) linearly and are (almost) equally spaced, already for 
comparatively weak interaction. 
The interaction strength at which this transition to the (almost) linear dependence occurs is proportional
to $1/(N-1)$ (see text).
In this graph the critical mean-field 
interaction strength is $c\approx 0.7$.
}
\label{fig8}
\end{figure}

\begin{figure}
\begin{center}
\centering
\subfigure{\epsfig{figure=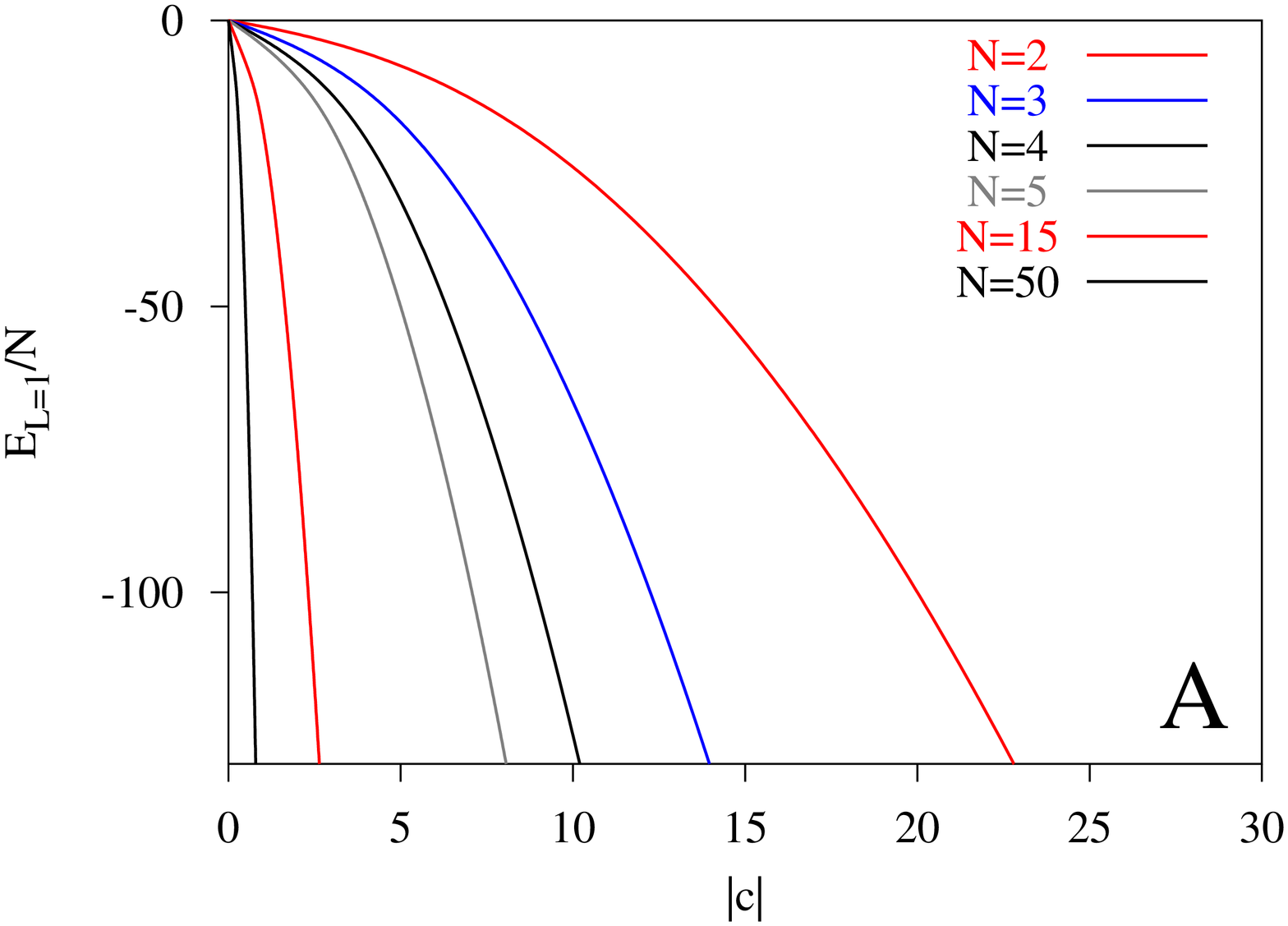,width=8cm, angle=-90}}
\subfigure{\epsfig{figure=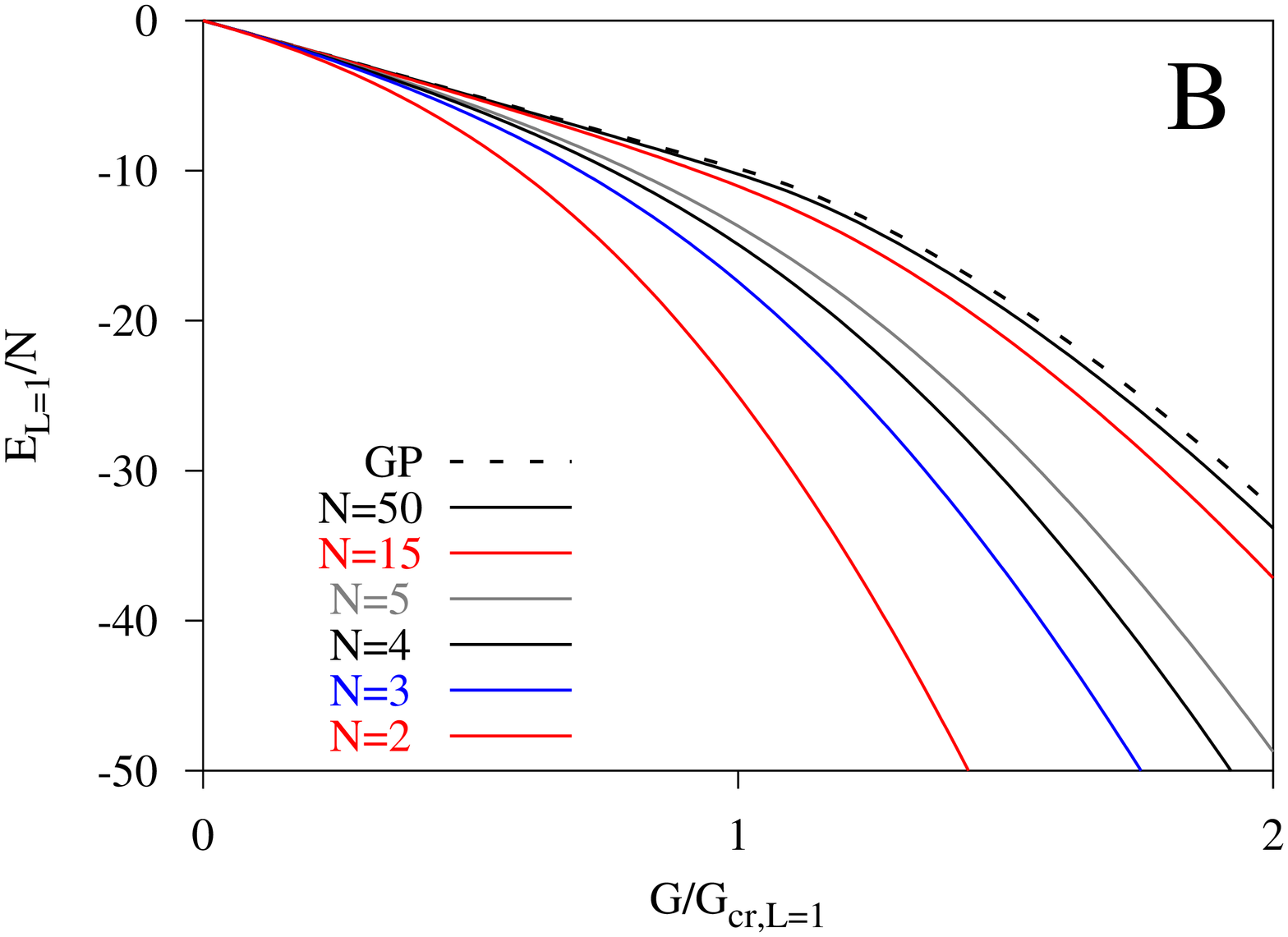,width=8cm, angle=-90}}
\end{center}
\caption{(color online) Energies per particle for attractive interaction.
The system resembles the system of $N$ attractive bosons on an infinite line.
This can be explained in a mean-field picture (see text).
A: Energies per particle as a function of $|c|$. With increasing particle numbers the energy per particle drops dramatically. 
B: Energies per particle as a function of $G/G_{cr,L}$, where $G=2c(N-1)/(2\pi)$ and $G_{cr,L}=-\pi/L$ is the critical and fixed
mean-field interaction strength. The GP energy is also shown (dashed line). For large $N$ the GP energy approaches the exact energy.
Length of the ring: $L=1$}
\label{fig9}
\end{figure}

In this section we discuss the results obtained by solving the coupled equations (\ref{DEFINING EQS})
for $c<0$, as described in section \ref{NumericalSolution}. Again, the $k_j$ are shown for fifteen particles in Fig. \ref{fig8}.
Since the $\delta_j$ are purely imaginary, only the imaginary part of the $k_j$ is plotted.
Similar to the repulsive case, the $k_j$ start to spread when $|c|$ is increased. However, 
for attractive interaction there is no saturation 
when $|c|\rightarrow\infty$ and the $k_j$ keep on spreading.
As can be seen from Fig. \ref{fig2}, the $\alpha_j$ and therefore also the $k_j$ grow virtually linear with $|c|$, 
already for comparatively weak attractive coupling, implying a quadratic dependence of the energy on $c$. 
This is the region in which the $\alpha_j$ are practically degenerate.
The larger the number of particles, the \emph{earlier} the degeneracy of the $\alpha_j$ sets in.
Fig. \ref{fig9}A shows the energy per particle for different particle numbers as a function of
$|c|$ on a ring of unit length. A linear dependence in the vicinity of $c=0$ is followed by a quadratic decrease for 
stronger interaction. Consistent with the perturbation theory results (\ref{1storderpert}), the
energy per particle for larger particle numbers is below that for smaller particle numbers.

\subsection{The relation to the system on an infinite line and to mean-field theory}
While in the repulsive case it is possible to compare the energy to either the TG limit 
or to the thermodynamic limit, neither of these exist in the attractive case. 
This follows from the behaviour of the exact solution, as discussed below. 
For attractive interaction we found that a different system is closely
related to the $N$ bosons on a finite ring.
Namely, this is the system of $N$ bosons subject to attractive $\delta$-function
interaction on an infinite line. This system has been solved exactly, see \cite{Calogero} and references therein.
The energy per particle is given by
\begin{equation}
\frac{E_{L=\infty}}{N}=-\frac{1}{12} c^2 (N^2-1).
\label{CalogeroEnergy}
\end{equation}
Although no periodic boundary conditions are imposed in this case, it is possible to think of this system as the 
limit $L\rightarrow\infty$ of the system on a ring of length $L$. However, the relation between these two systems is far less evident 
than the relation of the repulsive system to the impenetrable boson gas and the thermodynamic limit solution.
Therefore, we will motivate this relation in a simple mean-field picture before we begin with the comparison.
We briefly review the main mean-field results for the attractive case.

In the case of Bose-Einstein condensates the standard mean-field approximation is the so called Gross-Pitaevskii (GP) equation.
The GP equation has been solved analytically for the problem that we consider, namely 
the one-dimensional condensate on a ring \cite{Reinhardtrep,Reinhardtattr,Ueda}. 
Since the interaction strength appears in the GP equation only in combination with the particle number, 
the interaction strength can be parameterized
by the new parameter
\begin{equation}
G=\frac{2c(N-1)}{2\pi}.
\end{equation}
Thus, increasing the particle number or the interaction strength are equivalent in this approximation.
The ground state wave function is simply a constant as long as $|G|$ is less or equal to a critical $|G_{cr,L}|$, 
where $G_{cr,L}=-\frac{\pi}{L}$ for a ring of length $L$.
It is an angular momentum eigenfunction with zero angular momentum. 
As soon as $G/G_{cr,L}>1$ a second solution appears which is lower in energy than the constant solution, 
but is not an angular momentum eigenfunction, since it localizes at some arbitrary point on the ring.
For very strong attractive interaction it is essentially zero everywhere, 
except for the position around which it started to localize, 
much like a $\delta$-function. 
In this mean-field picture it is clear that the length of the ring loses its importance
in the strong interaction limit, since the wave function is localized on a small fraction of the ring and hence does not 
"see" the finite size of the ring. It should therefore make no difference how large the ring is. 
Furthermore, this effect should become important when $G/G_{cr,L}>1$.
Of course, this mean-field picture is oversimplifying, since the true
many-body wave function must be an eigenfunction of the angular momentum operator, 
but it captures some aspects of the true situation, as we shall show below. 

We return to  the discussion of the results.
Fig. \ref{fig9}B shows the energies per particle for different 
particle numbers as a function of the parameter $G/G_{cr,L}$ for $L=1$.
Since $G$ contains a factor of $N-1$, for any constant $G$ the energies per particle
for larger $N$ are now \emph{above} that for smaller $N$.
Also the GP ground state energy is depicted. 

\begin{figure}
\begin{center}
\centering
\subfigure{\epsfig{figure=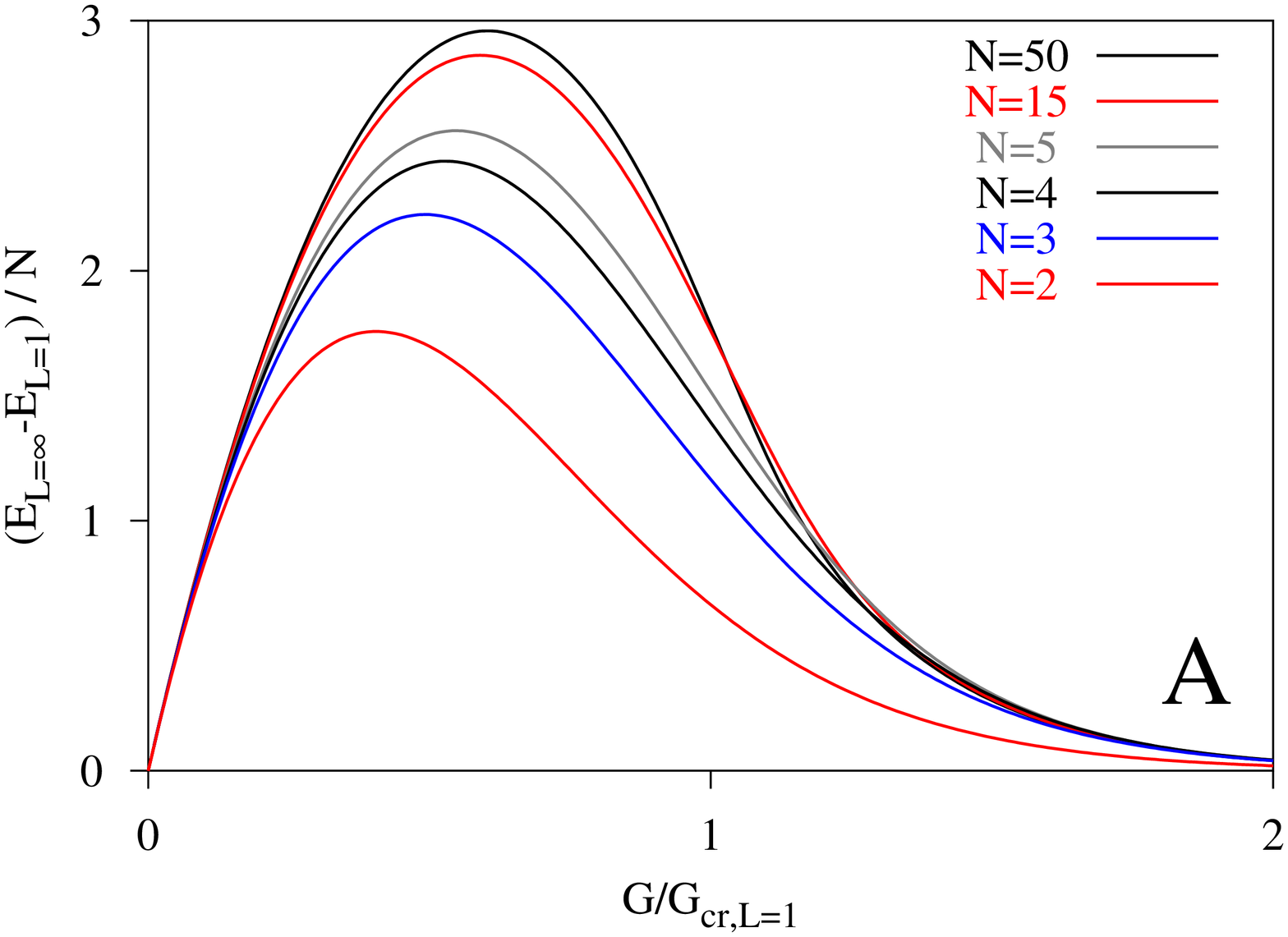,width=6cm, angle=-90}}
\subfigure{\epsfig{figure=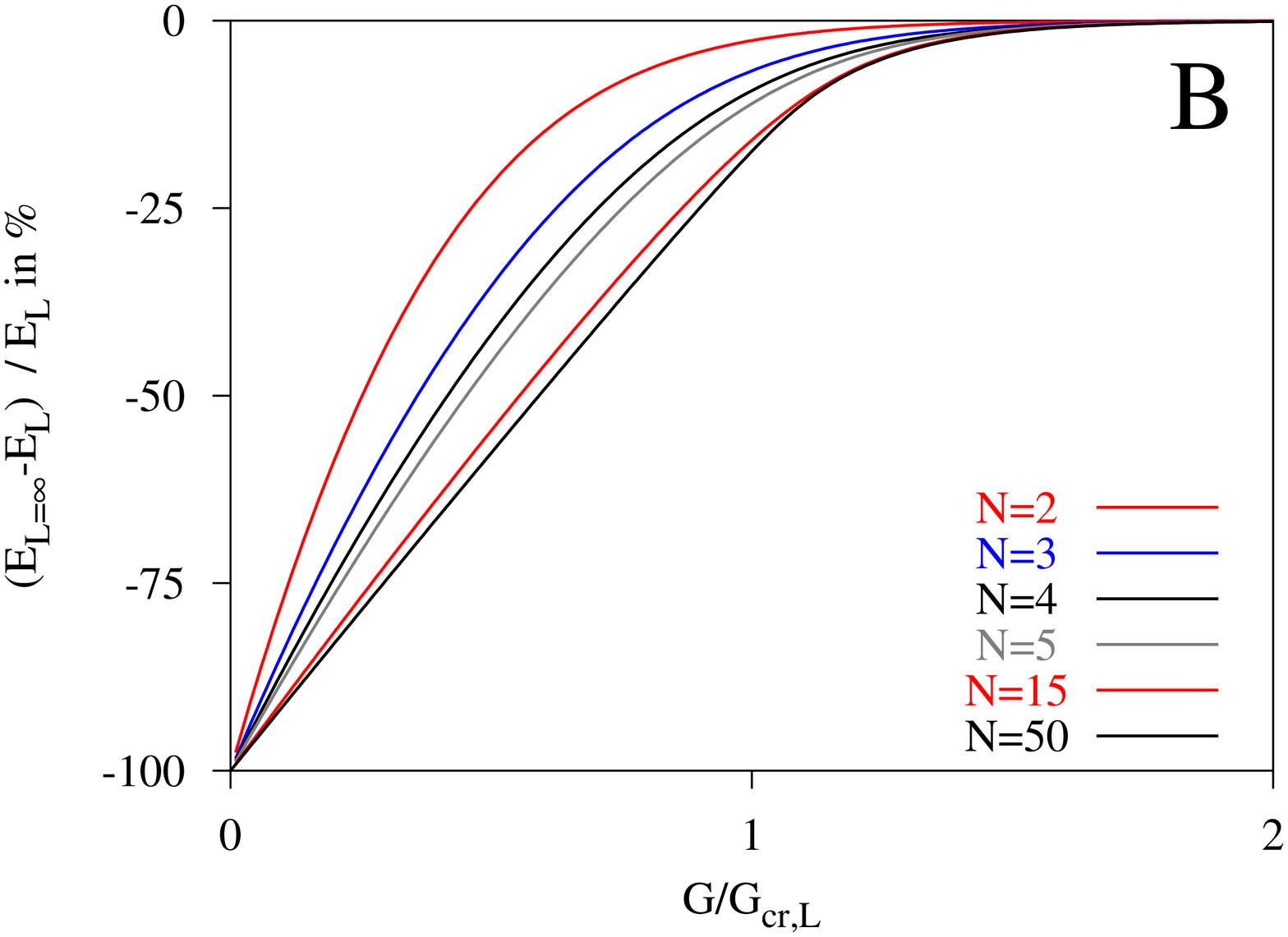,width=6cm, angle=-90}}
\end{center}
\caption{(color online) On the ability of $N$ attractive bosons on an infinite line to approximate the system of $N$ bosons on a ring.
A: Shown is the difference of the exact energies per particle $(E_{L=\infty}-E_L)/N$ of $N$ attractive bosons
on an infinite line and on a finite ring ($L=1$) as a function of $G/G_{cr,L}$.
This energy difference is counter intuitively \emph{positive}, implying that the energy on the ring of 
finite size is below that of the system on an infinite line.
It starts to decay to zero for
all particle numbers at approximately the same value of $G/G_{cr,L}$.
B: Shown is the relative error which is introduced by using the energy of the system of 
$N$ attractive bosons on an infinite line $L=\infty$ instead of the energy of the system on a ring of finite size.
As a function of $G/G_{cr,L}$ this graph is the same for all values of $L$.
This relative error decreases linearly for weak interaction, followed by an exponentially decaying tail.
}
\label{fig10}
\end{figure}

Now, we consider the difference between the exact energies per particle
on an infinite line and on a finite ring $(E_{L=\infty}(c)-E_{L}(c))/N$.
If the mean-field picture discussed above is correct, 
a sharp decrease in this energy difference is to be expected when the GP solution 
starts to localize. This should take place at $G=G_{cr,L}$ and therefore the corresponding critical 
$c$ values $c_{cr}=\frac{-\pi^2}{L(N-1)}$ should be proportional to $1/(N-1)$.
Fig. \ref{fig10}A shows that this is indeed the case. 
For all particle numbers the energy difference 
takes on a maximum, just before $G=G_{cr,L}$ and decreases rapidly afterwards.

We now consider  the relative and not the absolute deviation of the energy of 
$N$ attractive bosons on an infinite line from the energy of
those on a ring of length $L$. This deviation is given by 
\begin{equation}
\left(\frac{\Delta E_L}{E_L}\right)_{\infty}(c)=\frac{E_{L=\infty}(c)-E_{L}(c)}{E_{L}(c)}.
\label{errorLinfty}
\end{equation}
Similar to the limits to which we compared in the repulsive case, 
this is the error which is introduced by approximating the ring by a line of infinite length.
First, we discuss its scaling behaviour when $L$ varies and then its 
properties when considered as a function of $G/G_{cr,L}$ 
rather than $c$.
Using (\ref{scaleEnergy}) and the quadratic dependence of (\ref{CalogeroEnergy}) on $c$ one finds
that the relative deviation on a
ring of length $L$ is related to the relative deviation on a ring of length $L^{\prime}$ via the equation
\begin{equation}
\left(\frac{\Delta E_{L^{\prime}}}{E_{L^{\prime}}}\right)_{\infty}(c)=\left(\frac{\Delta E_L}{E_L}\right)_{\infty}(\frac{L^{\prime}}{L}c).
\label{scaleinferror}
\end{equation}
The relative deviation $\left(\frac{\Delta E_L}{E_L}\right)_{\infty}(c)$ for attractive
interaction scales exactly in the same way under a change of $L$ as the relative deviations
$\left(\frac{\Delta E}{E}\right)_{L,TDL}(c)$ and $\left(\frac{\Delta E}{E}\right)_{L,TG}(c)$
for repulsive interaction, although the limits considered are completely different.
It follows from (\ref{scaleinferror}) that the relative deviation $\left(\frac{\Delta E_{L}}{E_{L}}\right)_{\infty}(c)$ for a ring of
length $L$ is simply the relative deviation on a ring of unit length, evaluated at $\tilde{c}=Lc$.

How are the relative deviations $\left(\frac{\Delta E_L}{E_L}\right)_{\infty}(c)$ and 
$\left(\frac{\Delta E_L^{\prime}}{E_L^{\prime}}\right)_{\infty}(c^{\prime})$ on two rings of length $L$ and $L^{\prime}$ 
related when $c$ and $c^{\prime}$ are chosen to satisfy 
$G/G_{cr,L}=G^{\prime}/G_{cr,L^{\prime}}$?
In this case the interaction strengths on both rings equal the same 
fraction of the critical mean-field interaction strength on each ring.
The relation $G/G_{cr,L}=G^{\prime}/G_{cr,L^{\prime}}$ is equivalent to $c^{\prime}=c\frac{L}{L^{\prime}}$ 
which implies $\left(\frac{\Delta E_L}{E_L}\right)_{\infty}(c)=\left(\frac{\Delta E_L^{\prime}}{E_L^{\prime}}\right)_{\infty}(c^{\prime})$ when 
substituted into (\ref{scaleinferror}). 
The relative deviation on rings of different 
lengths is exactly the same, if the interaction strengths are the same
fraction of the critical mean-field interaction strength on each ring.
As a function of $G/G_{cr,L}$, it is therefore independent of $L$.
This relative deviation is depicted in  Fig. \ref{fig10}B.  
Its absolute value decreases linearly for $G/G_{cr,L}<1$
and exponentially for $G/G_{cr,L}>1$.
It can be seen clearly that for interaction strengths that are about twice as strong as the critical mean-field interaction strength 
the energy of the system of $N$ bosons on a 
ring is given to a good approximation by the energy of the system on an infinite line. 
This is equally true for all particle numbers studied.
Moreover, the factor $1/L$ in $G_{cr,L}$ implies that the larger the ring is, the smaller $c$ has to be for (\ref{CalogeroEnergy}) 
to be a good approximation to the exact ground state energy of the finite ring.

The fact that $(E_{L=\infty}(c)-E_{L}(c))/N$ and the relative deviation $\left(\frac{\Delta E_{L}}{E_L}\right)_{\infty}(c)$ 
converge to zero proves that the simple mean-field picture of a localizing wave function 
reproduces at least some of the true physical behaviour.
Interestingly, $(E_{L=\infty}(c)-E_{L})(c)/N$ is greater than zero, implying that the ground state 
of $N$ attractive bosons on an infinitely long line lies 
\emph{above} the ground state energy of those on a ring of finite size. 
Clearly, this is against the physical intuition. Energy levels are normally 
lowered when the distance between the confining boundaries of a system is increased. 
However, since periodic boundary conditions are used, there are no confining walls and this
\emph{anomalous} behaviour must be attributed to the
periodic boundary conditions, as was already pointed out earlier \cite{LiebLiniger,MugaSnider}.
This anomalous behaviour is obviously present for any particle number.

The weak-interaction behaviour of $\left(\frac{\Delta E_{L}}{E_L}\right)_{\infty}(c)$
can also be found by using (\ref{1storderpert}),(\ref{CalogeroEnergy}) and (\ref{scaleinferror}):
\begin{equation}
\left(\frac{\Delta E_L}{E_L}\right)_{\infty}(c)\stackrel{c\rightarrow 0}{\longrightarrow}\frac{1}{12}\,|c\,|L\,(N+1)-1=\frac{\pi^2}{12}\,\frac{G}{G_{cr,L}}\,\frac{N+1}{N-1}-1.
\label{EdiffLinfty}
\end{equation}
It can be seen in Fig. \ref{fig10}B that
the perturbative expression (\ref{EdiffLinfty}) 
gives a good approximation to $\left(\frac{\Delta E_L}{E_L}\right)_{\infty}(c)$ for large particle numbers
in the whole region $G/G_{cr,L}<1$, whereas for small particle numbers the 
validity of the approximation is very limited. 
This can be understood by remembering that the $c$ values corresponding to $G_{cr,L}$
are proportional to $N-1$. For large $N$ the critical $c$ are sufficiently close to zero and 
perturbation theory becomes applicable.
Therefore, only for large $N$ the expression (\ref{EdiffLinfty}) 
gives a good approximation to $\left(\frac{\Delta E_L}{E_L}\right)_{\infty}(c)$ 
in the whole region $G/G_{cr,L}<1$.

To conclude the discussion in the mean-field picture, it has to be mentioned that on an infinite line not only the 
exact ground state solution, but also the GP solution is known analytically.
Their energies are related by \cite{Calogero}
\begin{equation}
E^{(GP)}_{L=\infty}=E_{L=\infty}\, \frac{N}{N+1},
\label{CalogeroGP}
\end{equation}
implying that for $N\gg1$ and/or $G/G_{cr,L}\gg1$ even the GP energy $E^{(GP)}_{L=\infty}$ gives a good approximation to the 
exact energy of the finite system.
This is in contrast to the repulsive case, where the energy of the corresponding GP solution diverges for large $G$ and 
therefore does not reproduce the true physical behaviour, i.e. the TG limit \cite{Reinhardtrep}.
As mentioned earlier, the GP approximation unfortunately breaks the rotational symmetry of the Hamiltonian.
However, the situation can be remedied. Restoring the broken symmetry provides a 
ground state wave function of lower energy than the GP solution and unique 
many-body properties, see \cite{CCI}.

Since the energy of the system of $N$ bosons on an infinite line (\ref{CalogeroEnergy}) provides a good approximation 
to the energy per particle of the system on a finite ring, provided $|c|$ is not too small, 
it allows to investigate the possibility for different limits in the attractive case. 
The energy per particle diverges with $c^2$ for strong interaction and therefore there is no attractive equivalent 
to the TG limit.
Similarly, the thermodynamic limit does not exist since the energy per particle is always 
below (\ref{CalogeroEnergy}) which is independent of the size of the ring and
diverges with the number of particles as $N^2-1$.

\subsection{The importance of the finite length of the ring}
\begin{figure}
\includegraphics[width=11cm, angle=-90]{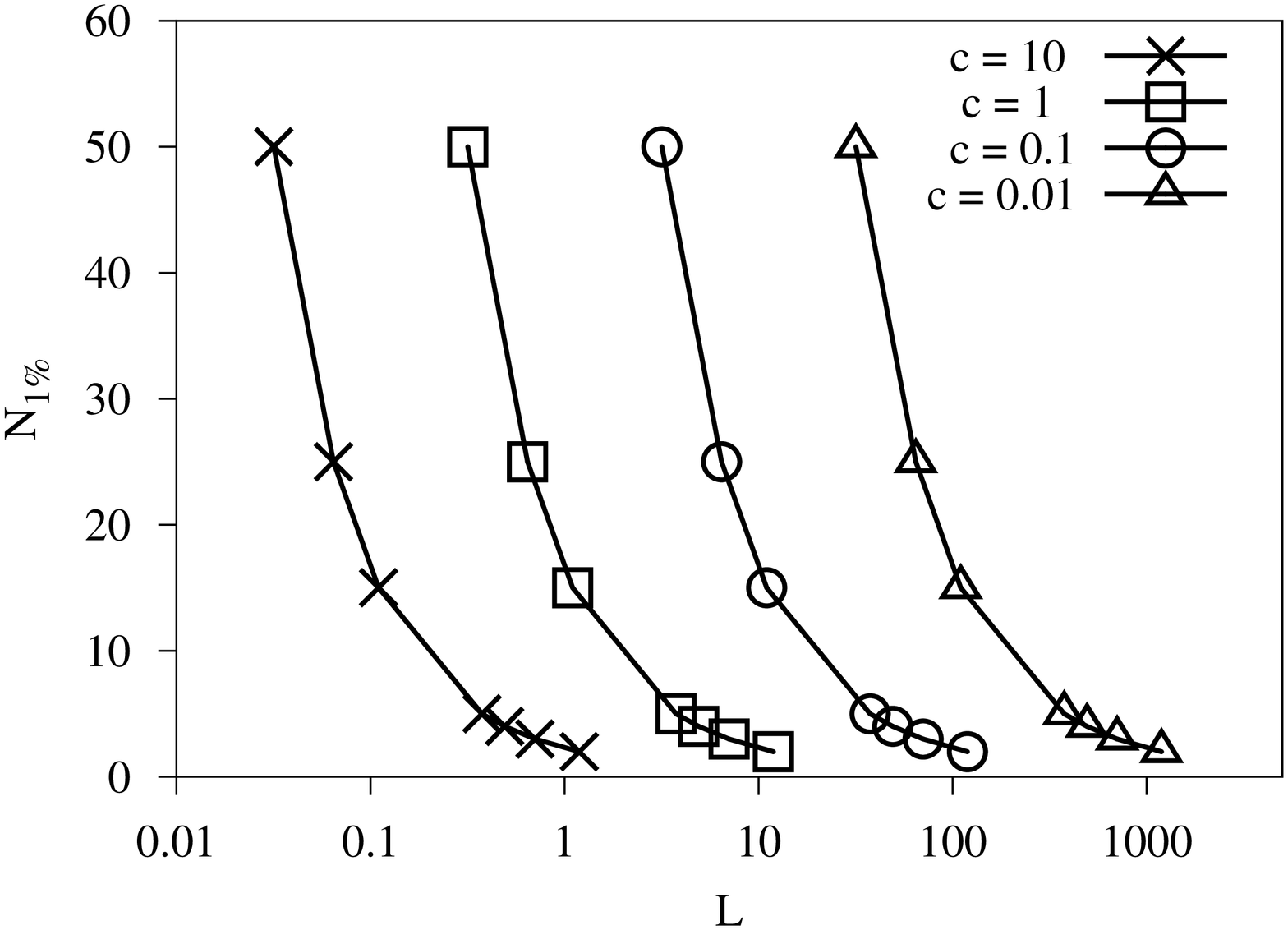}
\caption{
Depicted are the values of $N$ - denoted by $N_{1\%}$ - for which the 
error introduced by using the energy of $N$ attractive bosons on an infinite line instead of 
the energy of $N$ bosons on a ring of size $L$ is less than one percent.  
This error drops below one percent for all particle numbers
once $L$ is large enough. 
The curves can be shifted horizontally in the sense that a simultaneous change of $L\rightarrow x L$
and $c\rightarrow \frac{1}{x} c$ does not change $N_{1\%}$.
Note that the meaning of $N_{1\%}$ is different for repulsive and attractive interaction,
since the respective exact results are compared with different limiting situations.
}
\label{fig11}
\end{figure}

As we have shown above, the system of $N$ attractive bosons on a ring is closely related to the system of $N$ attractive bosons on an infinite line.
We found that the energies of these two systems are the closer, the larger the number of particles and the stronger the interaction is.
Now, we would like to address the following question: Given a ring of length $L$ and a fixed interaction strength $c$, 
how many particles are at least necessary that the energies of the two systems mentioned before differ by no more than one percent.
We denote this particle number by $N_{1\%}$, keeping in mind that $N_{1\%}$ has a different meaning for repulsive interactions.
Thus, we are looking for those particle numbers for which the relative deviation $\left(\frac{\Delta E_{L}}{E_L}\right)_{\infty}(c)$ defined in 
(\ref{errorLinfty}) is one percent. 
It is possible to think of $N_{1\%}$ as the particle number from which onwards the finite ring may be approximated by an infinite line within 
error bounds of one percent.
The result is shown in Fig. \ref{fig11}. For all particle numbers $\left(\frac{\Delta E_{L}}{E_L}\right)_{\infty}(c)$ drops below one percent,
if either $L$ or $c$ is large enough. This implies that it is possible for any particle number to approximate the finite ring by an infinite line
either by making the interaction stronger or the ring larger.
From (\ref{scaleinferror}) it follows that if one changes $L \rightarrow L^{\prime}$ and simultaneously
$c\rightarrow \frac{L}{L^{\prime}} c$, then the deviation $\left(\frac{\Delta E_{L}}{E_L}\right)_{\infty}(c)$ remains unchanged.
The curves can therefore be shifted horizontally in the sense that a change of $c\rightarrow \frac{1}{x}c$ is accompanied 
by a change $L\rightarrow xL$. 

The fact that the impact of the finiteness of the ring vanishes for strong interaction manifests itself also in the degeneracy of the
$\alpha_j$ which are depicted in Fig. \ref{fig2}B.
By considering the differences between the $\alpha_j$ as an effect which is due to the finite size of the ring, 
it is possible to derive an asymptotic relation between the energy of the system on an infinite line and that on a finite ring.
For any fixed value of $c$ it is possible to  choose a length of the ring $L_0$ such 
that $G/G_{cr,L}\gg 1$ for any $L \ge L_0$. 
The $\alpha_j$ are then virtually degenerate although they satisfy (\ref{order alpha_j}) at all times.
This can be seen in Fig. \ref{fig2}B, where $G/G_{cr,L}=1$ corresponds to $|c|\approx 0.7$.
To analyze this degeneracy we set $\alpha_j=\alpha=const$ for $j=1,\dots, N-1$  
All wave vectors $k_j$ are then equally spaced.
We calculate the energy of the system in this case, 
keeping in mind that for the ground state always $-k$ is in the
set $\{k_j\}$ when $k$ is.
By using (\ref{AA}) and (\ref{Alpha_j}) one finds that the energy 
(\ref{exact Energy}) for an odd number of particles is given by 
\begin{equation}
E=\sum_{j=1}^N k_j^2=2 \sum_{j=1}^{\frac{N-1}{2}}(-i\frac{\alpha}{L}\,j)^2=-\frac{1}{12}\left(\frac{\alpha}{L}\right)^2 N(N^2-1). 
\end{equation}
This is exactly the ground state energy of $N$ bosons on an infinite line (\ref{CalogeroEnergy}), if we put
\begin{equation} 
\alpha=|c|L. \label{AC}
\end{equation} 
An analogous calculation for even $N$ leads to the same expression.
We have thus found the asymptotic $c$-dependence of the ground-state $\alpha_j$ for $c \rightarrow - \infty $ and any particle number.
The result (\ref{AC}) for $N$ particles is consistent with the result of Muga and Snider for three particles \cite{MugaSnider}.
In principle, it is also possible to calculate a correction to this asymptotic $c$-dependence, for instance, 
by applying the techniques developed by Muga and Snider \cite{MugaSnider}, but we refrain from doing so due to the 
complexity of the equations (\ref{DEFINING EQS}) for large particle numbers.

\section{Conclusions \label{Conclusions}}
In this work we have presented the exact ground state solutions of the many-body Schr\"odinger equation for up to fifty bosons on 
a ring subject to pairwise $\delta$-function interaction. 
By employing the \emph{Bethe-ansatz} for 
attractive interactions we have proven numerically that the \emph{Bethe-ansatz} provides 
ground state solutions not only for repulsive interaction, where this is well known to be the case \cite{YangYang,Dorlas}, 
but also for attractive and all particle numbers studied.
We suspect that a complete set of states can be derived from a \emph{Bethe-ansatz} for \emph{any} particle number 
and \emph{any} interaction strength.

Our results show that the repulsive and the attractive case exhibit fundamentally different behaviour.
While in the repulsive case all wave vectors are purely real, they are purely imaginary in the attractive case, 
implying a jump in the derivative at the point of zero interaction strength.
For very strong repulsive interaction the wave vectors converge to \emph{finite} values which are equally spaced, 
in agreement with the result of Lieb and Liniger \cite{LiebLiniger}.
For strong attractive interaction on the other hand the wave vectors do not converge to finite values. 
We have derived an asymptotic relation for the wave vectors in this limit.
Asymptotically, the wave vectors for attractive interaction are then equally spaced, 
while their absolute distance keeps on growing with increasing interaction strength.

For repulsive interaction our exact data are compared in some detail with the energies 
determined in the thermodynamic limit and in the Tonks-Girardeau limit.
For attractive interaction, where these limits do not exist, we have compared the exact energies with those 
analytically known for a system on an infinite line. 

In the repulsive case we have shown that the energy of Lieb and Liniger's well known thermodynamic limit solution 
can differ substantially from that of our exact solution, 
especially when the number of particles is small and the interaction is \emph{weak}.
In detail, we found that the approximation of the 
finite system by the thermodynamic limit solution \emph{never} reproduces the exact energy 
within an accuracy of one percent, if the number of particles is less than eleven.
Moreover, our investigation has revealed that this thermodynamic limit approximation 
is accurate up to one percent for all repulsive interaction strengths only when the particle number is at least as large as $101$.
The results obtained allow one to conclude when the thermodynamic limit approximation is a good approximation and when 
finite $N$ effects have to be taken into account. This is not only of academic, but also of practical importance since in recent 
experiments on Bose-Einstein condensates only a few dozens of atoms could be studied, see \cite{expONED6} and references therein.
Hopefully, our results will be helpful to improve the description  of these experiments.

We have also investigated the system for finite $c$ in the light of the Tonks-Girardeau limit.
For strong repulsive interaction the energies of our exact solutions approach the Tonks-Girardeau energies. 
Our analysis proves that very strong repulsive interactions are necessary 
to approximate the exact energy by the Tonks-Girardeau energy within 
some well-defined error bounds. 
However, we found that the \emph{convergence} towards the Tonks-Girardeau limit 
is virtually independent of the number of particles when the particle density is held constant.
The relation revealed between the Tonks-Girardeau limit and our exact solution 
allows to estimate the exact ground state energy of the system  
for \emph{any} particle number and \emph{any} repulsive interaction strength.

In the attractive case we have related the system on a ring of finite size to that on an infinite line.
The ground state energies of these two systems are found to be essentially identical when the interaction is much stronger 
than a certain critical interaction strength. This critical interaction strength originates from mean-field theory and 
we could explain the strong relation between these two  systems in a mean-field picture.

Interestingly, the energy of the system on the finite ring is always below the corresponding energy of the system 
on an infinite line. This \emph{anomalous} behaviour was already found earlier for two and three particles \cite{LiebLiniger,MugaSnider} 
and was found to be present for any other particle number studied in this work.

For all interactions we have given explicit bounds on the minimal number of particles and the minimal 
size of the ring for the reference systems discussed to be good approximations to the exact solution.

Although the ground state for repulsive and attractive interaction is obtained by solving the same set of coupled transcendental 
equations the numerical effort for solving these equations differs substantially in the two cases.
While the system remains reasonably stable for repulsive interaction, the attractive case 
requires the use of very high precision numbers already for weak interactions.
  
Finally, we stress that the excitation spectrum of the system which is of high interest by itself
can also be obtained with the approach presented in this paper.

The explicit numbers of the exact ground state energies of the finite $N$ system studied in this work 
can be found on the Internet \cite{exactdata}.

%******************************
%The author acknowledges blabla
%******************************

%%%%%%%%%%%%%%%%%%%%%%%%%%%%%%%%%%%%%%%%%%%%%%%%%%%%%%%%%%%%%%%%%%%%%%%%%%%%%%%%%%%%%%%%%%%%%%%%%%%%

\pagebreak

\pagebreak

\pagebreak

\pagebreak

\pagebreak

\pagebreak

\pagebreak

\pagebreak

\pagebreak

%\pagebreak
%\begin{figure}
%\includegraphics[width=11.2cm, angle=-90]{./fig9.ps}
%\caption{Shown is the difference of the exact energies per particle $(E_{L=\infty}-E_L)/N$ between the system 
%of infinite length $L=\infty$ and that of finite length $L$ as a function of $G/G_{cr}$ for different $N$. 
%This difference is counterintuitively \emph{positive} and vanishes for 
%all particle numbers at approximately the same value $G/G_{cr}$.}
%\label{fig9}
%\end{figure}

\pagebreak

\pagebreak

\begin{thebibliography}{32}
\bibitem{expONED1}A. G\"orlitz et al.,  Phys. Rev. Lett. {\bf 87}, 130402 (2001).
\bibitem{expONED2}F. Schreck et al., Phys. Rev. Lett. {\bf 87}, 080403 (2001).
\bibitem{expONED3}M. Greiner, I. Bloch, O. Mandel, T. W. H\"ansch, and T. Esslinger, Phys. Rev. Lett. {\bf 87}, 160405 (2001).
\bibitem{expONED4}B. L. Tolra, K. M. O'Hara, J. H. Huckans, W. D. Phillips, S. L. Rolston, 
and J. V. Porto, Phys. Rev. Lett. {\bf 92}, 190401 (2004). 
\bibitem{expONED5}H. Moritz, T. St\"oferle, M. K\"ohl, and T. Esslinger, Phys. Rev. Lett. {\bf 91}, 250402 (2003).
\bibitem{expONED6}T. St\"oferle, H. Moritz, C. Schori, M. K\"ohl, and T. Esslinger, Phys. Rev. Lett. {\bf 92}, 130403 (2004).
\bibitem{expONED7}B. Paredes, A. Widera, V. Murg, O. Mandel, S. F\"olling, I. Cirac, G. V. Shlyapnikov, T. W. H\"ansch, and I. Bloch, Nature 429, 277.
\bibitem{Olshanii}M. Olshanii, Phys. Rev. Lett. {\bf 81}, 938 (1998).
\bibitem{GP1} E. P.\ Gross, Nuovo Cimento\  {\bf 20}, 454 (1961).
\bibitem{GP2} L. P.\ Pitaevskii, Sov.\ Phys.\ JETP {\bf 13}, 451 (1961).
\bibitem{Reinhardtrep}L. D. Carr, C. W. Clark, and W. P. Reinhardt, Phys Rev. A {\bf 62} 063610 (2000).
\bibitem{Reinhardtattr}L. D. Carr, C. W. Clark, and W. P. Reinhardt, Phys Rev. A {\bf 62} 063611 (2000).
\bibitem{Ueda}R. Kanamoto, H. Saito, and M. Ueda, Phys. Rev. A {\bf 67}, 013608 (2003).
\bibitem{Calogero} F. Calogero and A. Degasperis, Phys. Rev. A {\bf 11}, 265 (1975).
\bibitem{overview} A. Minguzzia, S. Succib, F. Toschib, M. P. Tosia, and P. Vignolo, Phys. Rep. {\bf 395}, 223 (2004). 
\bibitem{CCI} O. E.\ Alon, A. I.\ Streltsov, K.\ Sakmann, and L.S.\ Cederbaum, Europhys.\ Lett. {\bf 67}, 8 (2004).
\bibitem{CCI2} O. E. Alon, A. I. Streltsov, L. S. Cederbaum, Phys. Rev. B {\bf 71}, 125113 (2005).
\bibitem{BMF1} L. S.\ Cederbaum and A.I.\ Streltsov, Phys.\ Lett.\ A {\bf 318}, 564 (2003).
%\bibitem{BMF2} L.S.\ Cederbaum and A.I.\ Streltsov, Phys.\ Rev.\ A {\bf 70}, 023610 (2004).
%\bibitem{BMF3} A.I.\ Streltsov and L.S.\ Cederbaum, N. Moiseyev, Phys.\ Rev.\ A {\bf 70} 053607 (2004).
\bibitem{1Da} E. H. Lieb, \emph{Mathematical physics in one dimension}, Acad. Pr., New York, (1966). 
\bibitem{1Db} Z. Ha, \emph{Quantum many-body systems in one dimension}, World Scientific, Singapore, (1998).
\bibitem{LiebLiniger}E. H. Lieb and W. Liniger, Phys. Rev. {\bf 130}, 1605 (1963). 
\bibitem{Girardeau}M. Girardeau, J. Math. Phys. {\bf 1}, 516 (1960).
\bibitem{Tonks}L. Tonks Phys. Rev. {\bf 50}, 955 (1936).
\bibitem{pres1} G. E. Astrakharchik and S. Giorgini, Phys. Rev. A {\bf 68}, 031602(R) (2003).
\bibitem{pres2} M. D. Girardeau, Phys. Rev. Lett. {\bf 91}, 040401 (2003).
\bibitem{pres3} M. Olshanii and V. Dunjko, Phys. Rev. Lett. {\bf 91}, 090401 (2003).
\bibitem{pres4} E. H. Lieb, R. Seiringer, and J. Yngvason, Phys. Rev. Lett. {\bf 91}, 150401 (2003).
\bibitem{pres5} J. N. Fuchs, A. Recati, and W. Zwerger, Phys. Rev. Lett. {\bf 93}, 090408 (2004).
\bibitem{Lieb}E. H. Lieb, Phys. Rev. {\bf 130}, 1616 (1963).
\bibitem{cigar}V. Dunjko, V. Lorent, and M. Olshanii, Phys. Rev. Lett. {\bf 86}, 5413 (2001).
\bibitem{MugaSnider}J. G. Muga and R. F. Snider, Phys. Rev. A {\bf 57}, 3317 (1998).
\bibitem{YangYang}C. N. Yang and C. P. Yang, J. Math. Phys. {\bf 10}, 1115 (1969).
\bibitem{Dorlas}T. C. Dorlas, Commun. Math. Phys. {\bf 154}, 347 (1993).
%\bibitem{Rev2} F.\ Dalfovo, S.\  Giorgini, L.P.\ Pitaevskii, and S.\  Stringari, Rev.\ Mod.\ Phys.\ {\bf 71}, 463 (1999).
%\bibitem{inf1}F. A. Berezin, G. P. Pochil, V. M. Finkelberg, Moscow Univ. Vestnik {\bf 1}, 21 (1964).
%\bibitem{inf2}J. B. McGuire, J. Math. Phys. {\bf 5}, 622 (1964).
%\bibitem{inf3}E. Brezin and J. Zinn-Justin, C. R. Acad. Sci (Paris), {\bf B263}, 670 (1966).
%\bibitem{inf4}C. N. Yang, Phys. Rev. Lett. {\bf 19}, 1312 (1967) and Phys. Rev. {\bf 168}, 1920 (1968).
\bibitem{Bethe}H. A. Bethe, Z. Phys. {\bf 71}, 205 (1931).
\bibitem{Mathematica} Mathematica 5.0.1.0, Wolfram Research.
\bibitem{Olshaniiwebdata} http://physics.usc.edu/$\tilde{}$olshanii/DIST/1D\_gases/e\_f.dat.
%\bibitem{genLL1}Kunal K. Das, M. D. Girardeau and E. M. Wright, Phys. Rev. Lett. {\bf89}, 110402 (2002).
%\bibitem{genLL2}L. Salasnich, A. Parola, and L. Reatto, Phys. Rev. A {\bf 70}, 013606 (2004).
\bibitem{exactdata} http://www.pci.uni-heidelberg.de/tc/usr/kaspar/ring.html.




%\bibitem{Forrester}P. J. Forrester, N. E. Frankel, T. M. Garoni, N. S. Witte, Phys. Rev. A {\bf 67} 043607 (2003)

%%%%%%%%%%%%%%%%%%%%%%%%%%%%%%%%%%%%%%%%%%%%%%%%%%%%%%%%%%%%%%%%%%%%%%%%%%%%%%%%%%%%%%%%%%%%%%%%%%%%

\end{thebibliography}
\end{document}